\documentclass[printer]{aa}  

\usepackage{graphicx}
\usepackage{txfonts}
\usepackage{bm}
\usepackage{verbatim}
\usepackage[normalem]{ulem}
\usepackage{colortbl}
\usepackage{arydshln}
\usepackage{xcolor}

\def\dd{{\mathrm{d}}}
\def\LB{{L_\mathrm{B}}}
\def\mas{\mathrm{mas}}
\def\muas{\mu\mathrm{as}}
\def\yr{{\mathrm{yr}}}
\def\da{{\mathrm{day}}} 

\begin{document} 

   \title{Relativistic contributions to Mars rotation}


   \author{R.-M. Baland \inst{1}
          \and A. Hees \inst{2}
          \and M. Yseboodt \inst{1}
          \and A. Bourgoin \inst{2}
          \and S. Le Maistre \inst{1,3}
          }

   \institute{Royal Observatory of Belgium, Brussels, Belgium \\
\email{rose-marie.baland@oma.be}
   \and
   SYRTE, Observatoire de Paris, Université PSL, CNRS, Sorbonne Université, \\
   LNE, 61 avenue de l’Observatoire 75014 Paris, France \\
   \email{aurelien.hees@obspm.fr}
  \and
  UCLouvain, Louvain-la-Neuve, Belgium
   }

   \date{Received June, 2022; accepted , 2022}

  \abstract
   {The orientation and rotation of Mars, which can be described by a set of Euler angles (longitude, obliquity, and rotation angles), is estimated from radioscience data (tracking of orbiters and landers) and is then used to infer Mars internal properties. The data are  analyzed using a modeling expressed within the Barycentric Celestial Reference System (BCRS). This modeling includes several relativistic contributions that need to be taken properly into account to avoid a misinterpretation of the data.}
   {We provide new and more accurate (to the $0.1$ mas level) estimations of the relativistic corrections to be included in the BCRS model of the orientation and rotation of Mars.}
   {There are two types of relativistic contributions in Mars rotation and orientation: (i) those that directly impact the Euler angles and (ii) those resulting from the time transformation between a local Mars reference frame and BCRS. The former correspond essentially to the geodetic effect, but also to the smaller Lense-Thirring and Thomas precession effects. We compute them assuming that Mars evolves on a Keplerian orbit. As for the latter, we compute the effect of the time transformation and compare the rotation angle corrections obtained assuming that the planets evolve on Keplerian orbits with that obtained with realistic orbits as described by ephemerides. }
   {The relativistic correction in longitude comes mainly from the geodetic effect and results in the geodetic precession ($6.754\ \mas\ \yr^{-1}$) and the geodetic annual nutation ($0.565\ \mas$ amplitude). For the rotation angle, the correction is dominated by the effect of the time transformation. The main annual, semi-annual, and ter-annual terms have amplitudes of $166.954 \ \mas$, $7.783\ \mas$, and $0.544\ \mas$, respectively. The amplitude of the annual term differs by about $9\ \mas$ from the estimate usually considered by the community. We identify new terms at the Mars-Jupiter and Mars-Saturn synodic periods ($0.567\ \mas$ and $0.102\ \mas$ amplitude) that are relevant considering the current level of uncertainty of the measurements, as well as a contribution to the rotation rate ($7.3088\ \mas\ \da^{-1}$). There is no significant correction that applies to the obliquity.}
   {}

   \keywords{astrometry -- Planets and satellites: Mars --
                Relativistic processes --
                Reference systems}
\maketitle




\section{Introduction}

Apart from the Earth and the Moon, the rotation of Solar System bodies is inferred from data that are analyzed using the Barycentric Celestial Reference System (BCRS). Most of the time, the inferred rotation model is then used to estimate physical properties of these bodies related to their atmosphere/surface dynamics, or to their interior structure and composition. To prevent errors in the physical interpretation of the rotation models, one must properly correct them from relativistic contributions. 

Because many missions visited and studied Mars during the last decades, the rotation of the red planet has been thoroughly investigated. Many solutions have been proposed in the literature, based on different sets of data (mostly radiometric data from orbiters and/or landed spacecraft, e.g.~\citealp{Kono20, Kahan}). Along with these rotation models, interior models and Global Circulation Models (GCM) were also produced and improved over time, especially after the InSight mission. This is why we focus the present paper on Mars, although levels of relativistic contribution in the rotation of other planets are also provided at the end of this paper.

\begin{figure}[!htb]
\centering
\includegraphics[width=0.5\textwidth]{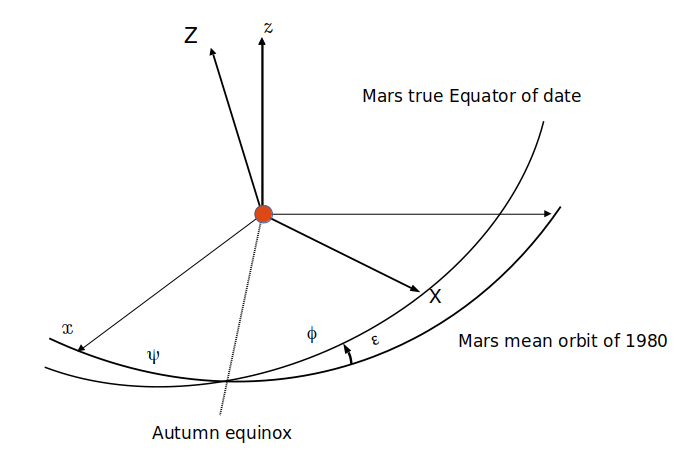}
\caption{Euler angles between the rotating Body Frame of Mars (axes $XYZ$) and the Inertial Frame associated with the mean orbit of Mars of 1980 (axes $xyz$, the $x-$axis is in the direction of the ascending node of Mars orbit over the Earth ecliptic of epoch J2000).
The $X-$axis of the BF is chosen as the prime meridian defined in the IAU convention \citep{IAU18}. 
The spin axis longitude $\psi$ is measured from the $x-$axis to the autumn equinox, $\phi$ is measured from the equinox to the $X-$axis, and the obliquity $\varepsilon$ is the angle from the $z-$axis to the $Z-$axis, or the inclination of the BF equator over the IF $xy$~plane. }\label{Fig1}
\end{figure}

The orientation of Mars with respect to its orbit can be described with a set of Euler angles (see Fig.~\ref{Fig1}), which are used to build the rotation models in the BCRS. These models are commonly corrected for relativistic effects in the longitude angle ($\psi$) and in the spin angle ($\phi$), while no relativistic correction in the obliquity ($\varepsilon$) of Mars is usually applied in the literature. 
The relativistic correction in $\psi$ is
oftenly not explicitly appearing in the angle definition\footnote{Except e.g.~in \citet{Kono11} or \citet{RK79} where the relativistic correction is written separately from the terms of Eq.~\eqref{Eqpsi}.}, but 
distributed between the different terms used to express that angle (e.g.~\citealp{Kono06}, Eq.~(14), \citealp{Folkner97b}, Eq.~(5), or \citealp{BMAN20}, Eq. (14)):
\begin{equation}
\psi(t) = \psi_0 + \dot\psi_0\, t + \psi_{nut}(t)\, ,
\label{Eqpsi} 
\end{equation}
where $\psi_0$ is the constant value at epoch J2000, $\dot\psi_0$ is the constant Mars precession rate, and $\psi_{nut}(t)$ is a periodic series of nutations. \citet{BMAN20} find a geodetic precession rate of $6.7$ mas/y hidden in $\dot\psi_0$ (see their Eq.~(69)), larger than the uncertainty on the determination of the precession rate ($\sim 2\, \mas$/y, \citealp{RisePaper}), and periodic geodetic nutations in $\psi_{nut}$, including an annual term with a $0.6$ mas amplitude (see also \citealp{Eroshkin2007}). 

As the Earth, Mars experiences periodic variations in its rotation due to atmosphere/surface dynamics that result in length-of-day (LOD) variations. By convention, the rotation angle of Mars $\phi$ is measured from the ascending node of the Mars true equator of date over the Mars mean orbit of epoch (usually chosen as the 1980 orbit , following \citealp{Folkner97}) to the intersection of Mars Prime Meridian on the Mars true equator of date (see Fig.~\ref{Fig1}). It is generally decomposed as (e.g.~\citealp{Kono06}, Eq.~(16))
\begin{eqnarray}
\nonumber \phi(t) &=& \phi_0 + \dot\phi_0\, t - \psi_{nut}(t)\, \cos\varepsilon_0 \\
&&+ \sum_{j=1}^4 \Big(\phi_{cj} \cos j\,l'(t) + \phi_{sj} \sin j\,l'(t)\Big) + [\phi]_\mathrm{GR}(t)\, ,
\label{Eqphi} 
\end{eqnarray}
where $\phi_0$ is the constant value at epoch J2000, $\dot\phi_0$ is the constant Mars spin rate, and $\phi_{cj}$ and $\phi_{sj}$ are the amplitudes of the periodic variations induced by the seasonal atmosphere/surface dynamics, with $l'(t)$ the mean anomaly of Mars. $\psi_{nut}(t)$ is the periodic nutation in longitude of Eq.~(\ref{Eqpsi}), and $\varepsilon_0$ is the J2000 epoch Mars obliquity. $t$ is the Barycentric Dynamical Time (TDB), see Sect.~\ref{Sect21}. The term $-\psi_{nut}(t) \cos\varepsilon_0$ in Eq.~(\ref{Eqphi}) is a correction for the fact that the seasonal periodic variations are measured along the Mars mean equator of date and not along the true equator of date, whereas the true equator of date is nutating with respect to the mean equator of date. This nutation term in the rotation angle therefore includes the periodic relativistic correction in longitude. The last term $[\phi]_\mathrm{GR}(t)$ in Eq.~(\ref{Eqphi}) is another relativistic correction arising when changing the time scale from Mars' proper time to TDB. This term takes into account: 
(i) the time dilation effect due to the barycentric orbital velocity of Mars' center-of-mass, 
(ii) the Einstein gravitational redshift effect due to the change of Mars' altitude inside the gravitational potential of the Sun caused by the non-vanishing eccentricity.
This relativistic correction is of the same order of magnitude as the seasonal periodic terms, and must be estimated accurately.

\citet{YS97} provided an estimation for $[\phi]_\mathrm{GR}(t)$ (see their Eq.~21, with amplitudes in mas):
\begin{eqnarray}
\nonumber [\phi]_\mathrm{GR}(t) &\approx& \sum_{j=1}^3\phi_{rj} \sin j\, l'(t) = - 175.80 \sin l'(t)\\
&&- 8.20 \sin 2l'(t) - 0.60 \sin 3l'(t)\, , 
\label{EqYS97}
\end{eqnarray}
that is still in use nowadays (e.g.~in \citealp{Kono20}). 
This estimation is expressed as a sum of trigonometric terms, the arguments being harmonics of the Mars mean anomaly $l'(t)$, as for the seasonal terms. As we will show in Sect.~\ref{Section3}, this estimation is affected by an error of about $9$ mas on the periodic terms, larger than the current formal uncertainty on the determination of rotation periodic variations ($\sim 1$ mas, \citealt{RisePaper}), and lacks of a linear term that affect the rotation rate.

To ensure a correct interpretation of rotation variations measurements, we aim to estimate the relativistic contribution to the Euler angles at the $0.1$ mas level. We will update the estimation of the main terms and will also investigate for the existence of so far neglected terms with amplitude larger than the $0.1$ mas threshold. 

The paper is organised as follow. In Sect.~2, we present the theory for the geodetic and Lense-Thirring effects, as well as for the effect of the time transformation. The results for the geodetic and Lense-Thirring effects are also presented in Sect.~2, whereas the results for the effect of the time transformation are presented in a dedicated part (Sect.~3) as it requires more investigations. Sect.~3 presents solutions obtained first assuming the planets evolve on Keplerian orbits and then on realistic orbits as described by ephemerides. In Sect.~4, we discuss the signature of the relativistic effects on the Doppler signal of a Martian radioscience instrument. Sect.~5 briefly introduces an application to the other planets of the Solar System. A discussion and conclusion are given in Sect.~6.

\section{Theory}

There exist two different types of relativistic contributions that arise in the BCRS rotation model of Solar System bodies: (i) contributions that impact directly the rotation of the body and (ii) contributions arising from reference frame transformation. The first type of contributions concerns the ones that impacts the spin equation of motion like the geodetic precession and nutations \citep{fukushima91, Eroshkin2007, BMAN20} and the Lense-Thirring and Thomas precessions (see Sect.~\ref{SectRelPRec}). The second relativistic contributions come from the reference frame transformation between a local inertial frame that would be used to describe the local physics of the body and the BCRS used to analyze the data. The theory of reference frame transformation to first post-Newtonian order has been derived by \citet{brumberg:1989fk,kopejkin:1988gd,damour:1991tp,klioner:1993tn} and has been adopted in the IAU 2000 conventions, see \citet{soffel:2003bd}. Of prime importance for our purpose is the time transformation between a local reference frame and BCRS. In Sect.~\ref{Sect21}, we will present in details various contributions that arise in the time transformation and their impact in terms of Mars rotation model.

\subsection{Geodetic, Lense-Thirring, and Thomas precession effects}
\label{SectRelPRec}

Within general relativity framework, the evolution of a spinning body is given, at the first post-Newtonian approximation, by a simple precession relation (see e.g., \citealp{1970PhRvD...2.1428B,soffel:2003bd, 2014grav.book.....P}):
\begin{equation}
\frac{\dd\textbf{S}}{\dd t} = \bm{\Omega}\wedge\textbf{S} + \mathcal{O}(c^{-4})\,\cdot
\label{eq:dSdt}
\end{equation}
In this equation, $\textbf{S}$ denotes the spin angular momentum of Mars and
$\bm{\Omega}$ is the relativistic total precessional angular velocity which is here decomposed into three parts: $\bm{\Omega}=\bm{\Omega}_{\mathrm{so}}+\bm{\Omega}_{\mathrm{ss}}+\bm{\Omega}_{\mathrm{TP}}$. The term $\bm{\Omega}_{\mathrm{so}}$ is called the spin-orbit precessional angular velocity, $\bm{\Omega}_{\mathrm{ss}}$ is the spin-spin precessional angular velocity, and $\bm{\Omega}_{\mathrm{TP}}$ is called the angular velocity of the Thomas precession. The spin-orbit and spin-spin components are also called the ``geodetic precession'' and the ``Schiff precession'' (or similarly the ``Lense-Thirring precession''), respectively.

When applying Eq. \eqref{eq:dSdt} to the description of the spin variations of Mars, one can keep the contribution of the Sun only in the relations of the spin-orbit and spin-spin precessional angular velocities, while the contribution of Phobos alone can be retained in the relation of the Thomas precession angular velocity. Thus, the expressions of the angular velocities read as follows
\begin{subequations}
\label{eq:Omega_SO_SS}
\begin{align}
  \bm{\Omega}_{\mathrm{so}}&=\frac{GM_\odot}{2c^2\Vert\textbf{x}-\textbf{x}_\odot\Vert^3}\:(\textbf{x}-\textbf{x}_\odot)\wedge(3\textbf{v}-4\textbf{v}_\odot)\, ,
  \label{eq:Omega_SO} \\
  \bm{\Omega}_{\mathrm{ss}}&=\frac{GS_\odot}{c^2\Vert\textbf{x}-\textbf{x}_\odot\Vert^3}\:\left(\frac{3\big((\textbf{x}-\textbf{x}_{\odot})\cdot\hat{\textbf{e}}_\odot\big)(\textbf{x}-\textbf{x}_{\odot})}{\Vert\textbf{x}-\textbf{x}_\odot\Vert^2}-\hat{\textbf{e}}_\odot\right)\, ,
  \label{eq:Omega_SS}\\
  \bm{\Omega}_{\mathrm{TP}}&=-\frac{1}{2c^2}\,\mathbf{v}\wedge\mathbf{Q}\, ,
  \label{eq:Omega_TP}  
\end{align}
\end{subequations}
where $\textbf x$ and $\textbf x_\odot$ are the barycentric positions of Mars and of the Sun, respectively, and $\textbf v$ and $\textbf v_\odot$ their barycentric velocities. The vector $\mathbf{Q}$ is the non-geodesic acceleration whose expression can be derived from Eq. (6.30a) of \cite{damour:1991tp}; hereafter, we only consider the dominant Newtonian contribution. $M_\odot$ is the mass of the Sun. $S_\odot$ and $\hat{\textbf{e}}_\odot$ denote the magnitude and direction of the Sun's spin angular momentum, respectively. $c$ is the speed of light and $G$ is the universal gravitational constant. 

Hereafter, we write $\hat{\textbf{e}}_\odot=\cos\alpha_{\odot} \cos\delta_{\odot}\,\hat{\mathbf{e}}_x + \sin\alpha_{\odot} \cos\delta_{\odot}\,\hat{\mathbf{e}}_y + \sin\delta_{\odot}\,\hat{\mathbf{e}}_z
$ with $\alpha_\odot$ and $\delta_\odot$ the right ascension and declination of the direction of the Sun's spin axis, respectively. Both are assumed to be fixed in the inertial frame associated with the mean orbit of Mars, namely $(\hat{\textbf{e}}_x, \hat{\textbf{e}}_y, \hat{\textbf{e}}_z)$. We also consider that $M/M_\odot\ll 1$, so that Mars follows an heliocentric orbit, namely $\Vert\textbf{x}_\odot\Vert/\Vert\textbf{x}\Vert\ll 1$, $\Vert\textbf{v}_\odot\Vert/\Vert\textbf{v}\Vert\ll 1$. We denote by $\hat{\textbf{e}}_z$ the direction of the orbital angular momentum of Mars (assumed to be constant) and by $\hat{\textbf{e}}_x$ the direction of the ascending node of Mars' orbit in the ecliptic; $\hat{\textbf{e}}_y$ completes the triad such that $(\hat{\textbf{e}}_x,\hat{\textbf{e}}_y,\hat{\textbf{e}}_z)$ is a direct orthogonal basis. After decomposing Eqs.~\eqref{eq:Omega_SO} and \eqref{eq:Omega_SS} into this basis (the case of the Thomas precession is treated separately at the end of the section), we obtain relationships as follows: 
\begin{subequations}
\begin{align}
 \bm{\Omega}_{\mathrm{so}} &= \Omega_{\mathrm{so}}\,(1+e\cos f)^3\,\hat{\textbf{e}}_z = \Omega_{\mathrm{so}}^z\, \hat{\textbf{e}}_z\, ,
 \label{EqOmegaso} \\
 \nonumber \bm{\Omega}_{\mathrm{ss}} &= \Omega_{\mathrm{ss}}\, \Bigg(\frac{\cos\delta_\odot}{2}\bigg(\Big(3\cos(2l'-\alpha_\odot) + \cos\alpha_\odot\Big)\,\hat{\textbf{e}}_x \bigg.\Bigg.\\
 &\Bigg.\bigg.+ \Big(3\sin(2l'-\alpha_\odot) + \sin\alpha_\odot\Big)\,\hat{\textbf{e}}_y\bigg) - \sin\delta_\odot \,\hat{\textbf{e}}_z\Bigg)\, ,
 \label{EqOmegass}
\end{align}
\end{subequations}
with
\begin{subequations}
\begin{align}
\Omega_{\mathrm{so}} &= \frac{3(GM_\odot)^{3/2}}{2c^2a^{5/2}(1-e^2)^{5/2}}\, ,\\
\Omega_{\mathrm{ss}} &=\frac{GS_{\odot}}{c^2a^3}\, , 
\end{align}
\end{subequations}
where $a$ is the semi-major axis, $e$ is the eccentricity, and $f$ is the true anomaly of Mars. We have neglected $e$ in $\bm{\Omega}_{\mathrm{ss}}$, as the Lense-Thirring effect obtained for a circular orbit is already very small (3 to 4 orders of magnitude smaller than the geodetic effect, see below).

We now write the components of $\hat{\textbf{e}}_x$, $\hat{\textbf{e}}_y$, and $\hat{\textbf{e}}_z$ in the coordinates of a rotating frame attached to Mars $(\hat{\mathbf{e}}_X,\hat{\mathbf{e}}_Y,\hat{\mathbf{e}}_Z)$, and oriented with the Euler angles $(\psi,\varepsilon,\phi)$ (see Fig.~\ref{Fig1}):
\begin{subequations}
\begin{align}
\nonumber \hat{\textbf{e}}_x &= (\cos\psi \cos\phi - \sin\psi \cos\varepsilon \sin\phi)\,\hat{\mathbf{e}}_X\\
&-(\sin\psi \cos\varepsilon \cos\phi + \cos\psi \sin\phi)\,\hat{\mathbf{e}}_Y + \sin\psi \sin\varepsilon \,\hat{\mathbf{e}}_Z\, , \\
\nonumber \hat{\textbf{e}}_y &= (\cos\psi \cos\varepsilon \sin\phi +\sin\psi \cos\phi) \,\hat{\mathbf{e}}_X \\
&+ (\cos\psi \cos\varepsilon \cos\phi - \sin\psi \sin\phi )\,\hat{\mathbf{e}}_Y - \cos\psi \sin\varepsilon \, \hat{\mathbf{e}}_Z\, , \\
\hat{\textbf{e}}_z &= \sin\varepsilon \sin\phi \,\hat{\mathbf{e}}_X + \sin\varepsilon \cos\phi \,\hat{\mathbf{e}}_Y + \cos\varepsilon \,\hat{\mathbf{e}}_Z\, \cdot
\end{align}
\end{subequations}
The precessional angular velocities $\bm\Omega_{\mathrm{so}}$ and $\bm\Omega_{\mathrm{ss}}$ can also be written as function of the Euler angles such as
\begin{eqnarray}
\nonumber \bm{\Omega} &=& (\dot\varepsilon\cos\phi + \dot\psi \sin\varepsilon \sin\phi)\,\hat{\mathbf{e}}_X-(
 \dot\varepsilon\sin\phi - \dot\psi \sin\varepsilon \cos\phi)\,\hat{\mathbf{e}}_Y\\
 && +(
 \dot\phi + \dot\psi \cos\varepsilon)\,\hat{\mathbf{e}}_Z\, ,
 \label{EqOmega}
\end{eqnarray}
where a ``dot'' denotes a differentiation with respect to time. 

By equating Eq.~(\ref{EqOmega}) with Eq.~(\ref{EqOmegaso}), we obtain $\dot\varepsilon=\dot\phi=0$, so that only the longitude angle $\psi$ is affected by the geodetic effect ($\dot\psi=\Omega_{\mathrm{so}}^z\neq0$). We proceed to a change of variable (from $t$ to $f$):
\begin{equation}
\dd\psi_{\mathrm{so}}(f)=\Omega_{\mathrm{so}}\, (1+e\cos f)^3\, \left(\frac{\dd{t}}{\dd{f}}\right)\,\dd{f}\, , 
\label{eq:dpsiSO}
\end{equation}
with $\dd{t}/\dd{f}$ being given by $p^{3/2}(GM_{\odot})^{-1/2}(1+e\cos f)^{-2}$ for a Keplerian motion, where $p=a(1-e^2)$ the semi-latus rectum of Mars' orbit. After substituting the expression of $\dd t/\dd f$ into the right-hand side of Eq.~\eqref{eq:dpsiSO}, the integration is immediate (see also Eq.~3 of \citealp{fukushima91}) and leads to
\begin{equation}
\psi_{\mathrm{so}}(f)=\frac{3}{2(1-e^2)}\,\left(\frac{na}{c}\right)^2\,(f+e\sin f)\, \cdot
\end{equation}
After using the equation of the center (see e.g., \citealp{2000ssd..book.....M}) to express the true anomaly in term of the mean anomaly $l'$, the expression for the spin axis longitude of Mars is given by
\begin{align}
 \nonumber  \psi_{\mathrm{so}}(l') &= \frac{3}{2(1-e^2)}\left(\frac{na}{c}\right)^2\\
  &\left(l'+\sum_{k=1}^{+\infty} \Bigg(e\sqrt{1-e^2}\,\bigg(J_{k-1}(k e)-J_{k+1}(k e) +\frac{2J_k(k e)}{ke\!\sqrt{1-e^2}}\bigg)\Bigg.\right.
  \nonumber \\
  & \left.\Bigg.+2\sum_{m=1}^{+\infty}\frac{\left(1-\sqrt{1-e^2}\right)^m}{ke^m}\bigg(J_{k-m}(ke) +J_{k+m}(ke)\bigg)\Bigg)\sin(kl')\right)\, ,
\end{align}
where the $J_{k}(x)$ are the Bessel functions of first kind with $k$-index, and where $n$ is the Mars' mean motion which is given by Kepler third law of motion. The term which is directly proportional to Mars' mean anomaly describes a precession in longitude at a steady rate while the other periodic terms represent the nutations in longitude. After making use of numerical values given in Table~\ref{tab:mars_param}, we find the following estimate reported here at third-order in eccentricity
\begin{eqnarray}
 \nonumber \psi_{\mathrm{so}}(t) &=& 6.754\ \mas\ \yr^{-1}\times t + 0.565\ \mas\sin{l'}\\
 &&+ 0.039\ \mas\sin{2l'} + 0.004\ \mas \sin{3l'}\, \cdot
  \label{EQ13}
\end{eqnarray}
Since the geodetic precession and nutations are small, the toy model based on the assumption of an elliptic Keplerian orbit is accurate enough for our purpose. The precession and annual terms are above the $0.1$ mas threshold and must be included in a model for the longitude angle $\psi$, as done e.g.~in \citet{BMAN20}. The geodetic precession term is larger than the uncertainty on the determination of the precession rate ($-7598.3\pm 2.1\ \mas\ \yr^{-1}$, \citealp{RisePaper}) and is needed to avoid an error of about $0.1\%$ in the determination of the polar moment of inertia. The geodetic annual term, with its amplitude of $0.6$ mas, does not depend on the properties of Mars interior, and has to be removed from any determination of the annual nutation term before interpretation in terms, for instance, of core radius.

\begin{table*}
\caption{Parameter values used for computing the relativistic contributions to Mars BCRS rotation model, in the frame of analytical and toy model developments of Sect.~\ref{Sect21} and \ref{Section31}, where the orbit of the planets are assumed to be Keplerian. The subscripts $\odot, J$ and $S$ refers to the Sun, Jupiter, and Saturn, respectively. }
\label{tab:mars_param}
\centering
\begin{tabular}{l l l }
\hline \hline
Parameter & Value & Reference \\
\hline
$G M_{\odot}$ & $1.3271244 \times 10^{20}$ m$^3$ s$^{-2}$& \citet{simon:2013vz} \\
$GM_J$ & $1.2671 \times 10^{17}$ m$^3$ s$^{-2}$& ibid \\
$GM_S$ & $3.794 \times 10^{16}$ m$^3$ s$^{-2}$& ibid \\
$a$ & $2.27939\times 10^{11}$ m  & ibid  \\ 
$a_J$ & $7.78298\times 10^{11}$ m  & ibid  \\
$a_S$ & $1.42939\times 10^{12}$ m  & ibid  \\
$e$ & $0.09340$ & ibid, as $\sqrt{(k^0)^2+(h^0)^2}$ \\
$c$ & $299792458$ m s$^{-1}$ & ibid \\ 
$n$ & $1.058576\times 10^{-7}$ rad s$^{-1}$ & ibid \\
$n_J$ & $1.678489\times 10^{-8}$ rad s$^{-1}$ & ibid \\
$n_S$ & $6.759040\times 10^{-9}$ rad s$^{-1}$ & ibid \\
$\dot\phi_0$ & $ 350^\circ.891985339$ day$^{-1}$ & \citet{RisePaper} \\
$\varepsilon_0$ & $25^\circ.18940927$& \citet{RisePaper}\\
$\psi_0$ &$35^\circ.437639$& adapted from \citet{RisePaper}\\
$S_\odot$ & $1.909\times 10^{41}$ kg m$^2$ s$^{-1}$& \citet{pijpers:1998tn} \\
$\alpha_\odot$ &$304^\circ.414$&adapted from \citet{IAU18}\\
$\delta_\odot$ &$84^\circ.351$& adapted from \citet{IAU18}\\
\hline
\end{tabular}
\end{table*}

By equating Eq.~(\ref{EqOmega}) with Eq.~(\ref{EqOmegass}), we obtain the equation of motion for the Euler angles related to the Lense-Thirring effect:
\begin{subequations}
\begin{align}
\dot\psi_{\mathrm{ss}} &= \tfrac{1}{2} \Omega_{\mathrm{ss}} \, \Bigg(-2 \sin\delta_{\odot} \Bigg.\\
&\Bigg. + \cos\delta_{\odot} \cot\varepsilon \, \bigg(\sin(\alpha_{\odot}-\psi) + 3 \sin(2 l'-\alpha_{\odot}-\psi) \bigg)\Bigg)\, , \\
\dot\varepsilon_{\mathrm{ss}} &= \tfrac{1}{2} \Omega_{\mathrm{ss}} \cos\delta_{\odot} \, \bigg(\cos(\alpha_{\odot}-\psi) + 3 \cos(2 l'-\alpha_{\odot}-\psi) \bigg)\, , \\
\dot\phi_{\mathrm{ss}} &= -\tfrac{1}{2} \Omega_{\mathrm{ss}} \frac{\cos\delta_{\odot}}{\sin\varepsilon} \, \bigg(\sin(\alpha_{\odot}-\psi) + 3 \sin(2 l'-\alpha_{\odot}-\psi) \bigg)\, \cdot
\end{align}
\end{subequations}
After integration (we considered the angles $\psi$ and $\varepsilon$ as constant in the right-hand sides of the equations, and therefore denote them with a subscript ``$0$'' in the following), the solution for each angle will be the sum of a linear term and of a periodic term at the semi annual period. After making use of numerical values given in Table~\ref{tab:mars_param}, we find the following estimate 
\begin{subequations}
\begin{eqnarray}
\psi_{\mathrm{ss}}(t) &=& -0.0779\ \muas\ \yr^{-1}\times t\ \sin\delta_\odot\nonumber\\
&&+0.0390\ \muas\ \yr^{-1}\times t\ \cot\varepsilon_0 \cos\delta_\odot \sin(\alpha_\odot-\psi_0)\nonumber\\
&& -0.0175\ \muas\ \cos(2l'-\alpha_\odot-\psi_0)\cot\varepsilon_0 \cos\delta_\odot\, ,\nonumber\\
&=& -0.0857\ \muas\ \yr^{-1}\times t \nonumber\\ 
&&- 0.0037\ \muas \cos(2 l'-339^\circ.852)\, ,\\
\varepsilon_{\mathrm{ss}}(t) &=& 0.0390\ \muas\ \yr^{-1}\times t\ \cos\delta_\odot\cos(\alpha_\odot-\psi_0)\nonumber \\
&&+ 0.0175\ \muas\sin(2l'-\alpha_\odot-\psi_0)\cos\delta_\odot\, 
,\nonumber\\
&=&-0.0001 \ \muas\ \yr^{-1}\times t \nonumber\\ 
&&+ 0.0017\ \muas\sin(2 l'-339^\circ.852)\, ,\\
\phi_{\mathrm{ss}}(t) &=& -0.0390\ \muas\ \yr^{-1}\times t\ \frac{\cos\delta_\odot}{\sin\varepsilon_0}\sin(\alpha_\odot-\psi_0) \nonumber \\
&&+ 0.0175\ \muas\cos(2l'-\alpha_\odot-\psi_0)\ \frac{\cos\delta_\odot}{\sin\varepsilon_0}\, ,
\nonumber\\
&=& 0.0090 \ \muas\ \yr^{-1}\times t\nonumber\\ 
&&+0.0040\ \muas\cos(2 l'-339^\circ.852)\, \cdot
\end{eqnarray}
\end{subequations}
We actually see that the spin-spin contribution is well below the $0.1\ \mas$ precision and can therefore be safely neglected, compare to the spin-orbit contribution.

While considering the spin of Mars as an accelerated gyroscope, we shall also consider the Thomas precession $\mathbf{\Omega}_{\mathrm{TP}}$ contribution inside the total relativistic angular precessional velocity $\mathbf{\Omega}$ (see e.g., Eq.~(25) of \citealp{soffel:2003bd}). As shown in Eq. \eqref{eq:Omega_TP}, this term is proportionnal to the acceleration which characterises the deviation of the actual worldline of the planet from a geodesic, which comes mainly from the coupling of higher order multipole moments of Mars to the external tidal gravitational fields. The Thomas precession scales as $\Omega_{\mathrm{TP}}=\Vert\bm{\Omega}_{\mathrm{TP}}\Vert\propto Q\Vert\mathbf{v}\Vert/c^2$ with $Q=\Vert\mathbf{Q}\Vert\propto J_2R^2 GM_{p}/\Vert\mathbf{x}-\mathbf{x}_p\Vert^4$, where $J_2$ and $R$ are respectively the quadrupole moment and the equatorial radius of Mars while $M_p$ and $\mathbf{x}_p$ are respectively the mass and the barycentric position of the external body. For the Earth, the dominant contribution to Thomas precession comes from the Moon. Here for Mars, Phobos plays the dominant role. The absolute value of $Q$ due to the action of Phobos is on the order of $10^{-12}\ \mathrm{m}\ \mathrm{s}^{-2}$, meaning that $\Omega_{\mathrm{TP}}\sim 10^{-6}\ \muas\ \yr^{-1}$ (against $10^{5}\ \muas\ \yr^{-1}$ and $10^{-1}\ \muas\ \yr^{-1}$ for $\Omega_{\mathrm{so}}$ and $\Omega_{\mathrm{ss}}$, respectively). The contribution of Thomas precession to the variations in Euler angles is therefore smaller than the already negligible contribution of the Lense-Thirring precession.

\subsection{Time coordinate transformation and impact on Mars rotation modeling}
\label{Sect21}

We will denote by $\tau$ the local time (or proper time) related to the central body (i.e. Mars) and by $t$ the Barycentric Dynamical Time (TDB), related to the BCRS, and used to analyze Mars data. The local model of rotation describes the rotation of the body in terms of local physics like atmosphere/surface dynamics. This model is typically expressed as a function of the local time $\tau$. For example, for a uniformly rotating body $\phi=\phi_0 + \dot\phi_0 \tau$. As a consequence, when expressed in BCRS, the rotation model will be impacted by the $\tau\!-\!t$ time transformation. In Sect.~\ref{sec:time_transfo}, we present the theory related to the time transformation while in Sect.~\ref{sec:rot_BCRS}, we show how this time transformation impacts the modeling of Mars' rotation when expressed in BCRS.

\subsubsection{Time coordinate transformation}
\label{sec:time_transfo}

The Barycentric Dynamical Time $t$ is a rescaled version of Barycentric Coordinate Time (TCB).
The link between TDB and TCB is defined by the recommendation B3 of IAU2006 which reads
\begin{equation}
\frac{\dd t}{\dd \mathrm{TCB}} = 1 - \LB\, ,
\label{EqLB}
\end{equation}
where $\LB = 1.550519768\times10^{-8}$, see Eq.~(10.3) from \citet{petit:2010fk}. 

The proper time related to a local frame co-moving with Mars (the Martian equivalent to Geocentric Coordinate Time, or TCG) will be here denoted by $\tau$. For a small orbital velocity ($v \ll c$) and a weak gravitational field ($r \gg GM/c^2$), the relationship between TDB and Mars proper time $\tau$ at first post-Newtonian order is given by (see e.g.~\citealp{soffel:2003bd,petit:2010fk})
\begin{align}
\frac{\dd \tau}{\dd t} -1&=\left[\frac{\dd \tau}{\dd t}\right]_{GR}= \frac{\LB}{1-\LB}-\frac{1}{c^2}\frac{1}{1-\LB}\left(\frac{v^2}{2}+w\right)\nonumber \\
&+\frac{1}{c^4}\frac{1}{1-\LB}\left(-\frac{v^4}{8}-\frac{3}{2}v^2 w + 4\textbf v\cdot\textbf w + \frac{1}{2}w^2 + \Delta \right),\, 
\label{Eqtau2}
\end{align}
where $\textbf{v}$ is Mars' barycentric velocity. The potential $w$ is the Newtonian potential at Mars' location
\begin{equation}
  w = \left(\sum_{A\neq \mathrm{Mars}} \frac{GM_A}{r_A}\right) - \frac{3J_2^\odot}{2}\frac{GM_\odot}{r_\odot} \left(\frac{R_\odot}{r_\odot}\right)^2 \left(\left(\frac{\hat{\textbf{e}}_\odot\cdot (\textbf x-\textbf x_\odot)}{r_\odot}\right)^2-\frac{1}{3}\right)\, ,
\end{equation}
where $r_A=\Vert\textbf x-\textbf x_A\Vert$ with $\textbf x$ the barycentric position of Mars and $\textbf x_A$ the position of the body $A$ and where the sum includes the Sun and all planets, $R_\odot$ is the Sun's equatorial radius, $\hat{\textbf{e}}_\odot$ is the unit vector defining the Sun's spin axis and $J_2^\odot$ is the Sun's quadrupolar moment of the mass distribution. The norm of the barycentric distance between Mars and the Sun is denoted $r_\odot = \Vert\textbf{x} - \textbf{x}_\odot\Vert$. In addition, $\textbf w$ is the vector potential defined by
\begin{equation}
  \textbf w =\sum_{A\neq \mathrm{Mars}} \frac{GM_A}{r_A} \textbf v_A \, ,
\end{equation}
and $\Delta$ is defined by
\begin{eqnarray}
  \Delta &=& \sum_{A\neq \mathrm{Mars}} \frac{GM_A}{r_A}\Bigg(-2\  v_A^2 +\Bigg.\nonumber\\
  &&\Bigg. \sum_{B\neq A} \frac{GM_B}{R_{AB}}+\frac{1}{2}\left(\frac{\textbf v_A\cdot \left(\textbf x-\textbf x_A\right)}{r_A}\right)^2+\frac{1}{2}\ \textbf a_A\cdot (\textbf x-\textbf x_A)\Bigg)\, ,
\end{eqnarray}
where $\textbf a_A$ is the Newtonian point-mass acceleration of body $A$, namely
\begin{equation}
  \textbf a_A=\sum_{B\neq A}\frac{GM_B}{\Vert\textbf{x}_B-\textbf{x}_A\Vert^3}(\textbf{x}_B-\textbf{x}_A)\, ,
\end{equation}

We can write the result from the integration of Eq.~(\ref{Eqtau2}), the relationship between TDB and Mars’ proper time $\tau$, as
\begin{equation}
  \tau = t+[\tau\!-\!t]_\mathrm{GR}\, ,
\label{EqtauTDB}
\end{equation}
where $[\tau\!-\!t]_\mathrm{GR}$ includes the various relativistic corrections. Typically the relative amplitude of $[\tau\!-\!t]_\mathrm{GR}$ is of the order of $GM_\odot/c^2 a\sim 10^{-8}$ where $a$ is Mars' semi-major axis. In the following, we will show how these relativistic corrections impact the data analysis related to Mars rotation when analyzed using TDB.

\subsubsection{Impact on Mars rotation modeling}
\label{sec:rot_BCRS}
We now explore how the time transformation developed above can impact Mars rotation modeling in BCRS. Let us first consider a rotation model expressed in Mars' local reference frame. Such a model includes a uniform rotation and periodic terms, i.e.
\begin{equation}
 \phi_\mathrm{local}(\tau) = \phi_0 + \dot\phi_{0}^{\mathrm{local}} \, \tau - \psi_{nut}\, \cos\varepsilon_0 + \sum_{j=1}^4 (\phi_{cj}\cos j\,l' + \phi_{sj} \sin j\,l') \, .
\end{equation}
On the other hand, one needs a similar modeling expressed in terms of TDB in order to perform the data analysis. Using Eq.~(\ref{EqtauTDB}), this modeling is given by 
\begin{eqnarray}
  \phi(t) &=& \phi_\mathrm{local}\left(\tau(t)\right)
  = \phi_0 + \dot\phi_{0}^{\mathrm{local}} \, t - \psi_{nut}\, \cos\varepsilon_0\nonumber \\
  && + \sum_{j=1}^4 (\phi_{cj}\cos j\,l'+\phi_{sj}\sin j\,l') + \left[\phi\right]_\mathrm{GR}\!(t)\, ,
\label{eq:phi_t}
\end{eqnarray}
where
\begin{equation}
\left[\phi\right]_\mathrm{GR}\!(t) =
\dot\phi_{0}^{\mathrm{local}}\, \left[\tau\!-\!t\right]_\mathrm{GR} \, \cdot
\label{eq:phi_GR}
\end{equation}

When analysing data from orbiting spacecraft or surface landers, the time scale used in the data reduction is TDB, so that the model for Mars rotation needs to include the relativistic contributions as presented above.
In the following section, we will study numerically and analytically various contributions that impact the $[\tau\!-\!t]_\mathrm{GR}$ relationship and their impact on Mars' rotation angle $\phi$. We will propose a new modeling improving the one currently used in various analyses. 

Additional relativistic corrections in longitude and obliquity can be computed similarly as for the rotation angle, by replacing the rotation rate in Eq.~(\ref{eq:phi_GR}) by the precession rate in longitude and in obliquity. However, as those rates ($-7598.3\ \mas\ \yr^{-1}$ and $-7.9\ \mas\ \yr^{-1}$, respectively, see \citealp{RisePaper}) are very small compared to the rotation rate, the associated relativistic corrections are negligible.

\section{Impact of the time transformation on Mars rotation modeling}
\label{Section3}

In this section, we study the various contributions to the $\left[\tau\!-\!t\right]_\mathrm{GR}$ relationship that arises from integrating Eq.~(\ref{Eqtau2}) and their corresponding impacts on Mars' BCRS rotation model through Eq.~(\ref{eq:phi_GR}). In order to cross-check and validate our results, we develop three independent approaches. 

The first approach consists in developing a simple toy model to integrate analytically Eq.~(\ref{Eqtau2}). This has the advantage to provide a good physical intuition on the various terms obtained and is relatively pedagogical.

The second approach consists in integrating numerically Eq.~(\ref{Eqtau2}) using the DE440 planetary ephemerides \citep{park:2021wd} provided by the NAIF-SPICE software \citep{acton:1996fk,acton:2018aa}. In a second step, analytical series consisting in various harmonic terms are fitted to the result of the numerical integration. This procedure is similar to the one developed to produce the Time Ephemeris for Earth, see \citet{fukushima:1995uq, irwin:1999fk, harada:2003vv, fukushima:2010yu}.

The third approach we develop consists in using series for the barycentric position and distance of planets to obtain series for the relativistic part of the rotation angle. We use the analytical planetary theory VSOP87 \citep{VSOP87}, derived from the DE200 planetary ephemerides \citep{DE200}. Using the VSOP87 series, we analytically integrate Eq.~(\ref{Eqtau2}) and identify the harmonics with the largest contribution to the $\left[\tau\!-\!t\right]_\mathrm{GR}$ relationship. We purposely use VSOP87 and not the more recent versions of VSOP theories that are not suited for our purpose. VSOP2000 \citep{VSOP2000} provides series only for the heliocentric and not barycentric positions, whereas VSOP2013 \citep{simon:2013vz}, derived from the INPOP planetary ephemerides \citep{fienga:2011qf,bernus:2019aa,bernus:2022wc}, is based on Tchebychev polynomials and not on series. We will show that the error introduced by using the older, and therefore less accurate, VSOP87 theory is negligible for our purpose, as the solutions of the second and third approaches are consistent to the 0.1 mas level.

The advantages to consider the last two approaches rely in the fact that we can cross-check our results and estimate the uncertainties in our derived modeling coming from numerical integration, differences between the DE and INPOP ephemerides, etc. 

In the following of this section, we will consider the various contributions from Eq.~(\ref{Eqtau2}): the $1/c^2$ contribution coming from the 2-body problem, the contribution related to the motion of the Sun with respect to the Solar System Barycenter (SSB), the contribution from the $\LB$ constant, the direct contribution from other planet, and the higher order contributions (the $1/c^4$ terms and the Sun's $J_2^\odot$).

\subsection{Simple analytical solution (toy model)}
\label{Section31}

\subsubsection{The $1/c^2$ contribution coming from the Sun considering a Keplerian motion}
\label{Section311}
We first consider the main contribution to the relation between $\tau$ and TDB, which is the $1/c^2$ contribution from the Sun in a 2-body problem (or Keplerian problem). 

The evolution of proper time with respect to coordinate time is given by
\begin{equation}
    \frac{\dd \tau}{\dd t} -1=\left[\frac{d\tau}{d t}\right]_{\mathrm{2body}} = \frac{\LB}{1-\LB} - \frac{1}{1-\LB}\frac{v^2_{M\odot}}{2c^2} - \frac{1}{1-\LB}\frac{GM_\odot}{r_{M\odot}c^2}\, ,
\label{eq:dtau_dt_2body}
\end{equation}
where $r_{M\odot}$ is the distance between the Sun and Mars and $v_{M\odot}$ is the norm of their relative velocity. This expression can be integrated exactly assuming a perfect Keplerian motion, see e.g.~\citet{moyer:1981kx}:
\begin{eqnarray}
  \left[\tau\!-\!t\right]_{\mathrm{2body}} &=& \mathrm{cst} + \frac{\LB}{1-\LB}\, t - \frac{1}{1-\LB}\, \frac{n a^2}{2c^2}\,\left(4 E - n t \right)\, ,\nonumber\\
  &=& \mathrm{cst} + \frac{\LB}{1-\LB}t - \frac{1}{1-\LB} \frac{n a^2}{c^2}\left(2 e \sin E +\frac{3}{2} l' \right)\, ,
\end{eqnarray}
where $n$ is the mean motion, $a$ the semi-major axis, $E$ the eccentric anomaly and $l'$ the mean anomaly. A low eccentricity expansion leads to
\begin{eqnarray}
  \left[\tau\!-\!t\right]_{\mathrm{2body}} &=&  \mathrm{cst} + \frac{1}{1-\LB} \left(\LB-\frac{3n^2a^2}{2c^2}\right) t\nonumber \\
  &&  -\frac{na^2}{c^2(1-\LB)} \left(2e-\frac{e^3}{4}\right) \sin l' \nonumber \\
  && - \frac{na^2}{c^2(1-\LB)}\left(e^2-\frac{e^4}{3}\right) \sin 2l' \nonumber \\
  && - \frac{3na^2}{4c^2(1-\LB)}\, e^3 \sin 3l' \nonumber\\
  && - \frac{2na^2}{3c^2(1-\LB)}\, e^4 \sin 4l' + \dots \, ,
  \label{Eq27}
\end{eqnarray}
which includes a linear drift and oscillations at frequencies multiple of the orbital frequency. The linear drift includes a contribution from the rescaling between TCB and TDB (i.e. the contribution from $\LB$) of $1.55\times 10^{-8}$ and a contribution of $-9.72 \times 10^{-9}$ from the Sun (using Mars' orbital parameters from Table~\ref{tab:mars_param}). The total linear drift coefficient is $5.79\times 10^{-9}$. The amplitudes of the harmonics terms are: -11.419 ms for the term at orbital period, -532.3 $\mu$s for the term at twice the orbital period, -37.4 $\mu$s for the term at three times the orbital period and $-3.1$ $\mu$s for the term at four times the orbital period. Note that the $1/(1-\LB)$ coefficient impacts these amplitude only at the relative level of $10^{-8}$.

The impact on Mars rotation is obtained from Eq.~(\ref{eq:phi_t}). First of all, it is important to notice that the $\dot\phi_0$ estimated from a data analysis performed in BCRS is actually
\begin{equation}
  \dot\phi_0 = \dot\phi_{0}^{\mathrm{local}}\left(1+\frac{1}{1-\LB}\left(\LB - \frac{3n^2a^2}{2c^2}\right)\right) = \dot\phi_{0}^{\mathrm{local}} \left(1+5.79\times 10^{-9}\right) \, ,
\end{equation} 
where $\dot\phi_{0}^{\mathrm{local}}$ is the proper Mars' rotation rate. 
Using the measured value of $\dot\phi_0$ from Table~\ref{tab:mars_param}, one finds that
\begin{equation}
\dot\phi_{0}^{\mathrm{local}} = 350^\circ.891983308 \, \da^{-1} \, ,
\end{equation}
such that $\dot\phi_\mathrm{GR} = 5.79 \times 10^{-9} \, \dot\phi_{0}^{\mathrm{local}}$ is $2^\circ.03 \times 10^{-6} \, \da^{-1}$ (or $7.3117 \ \mas\ \da^{-1}$).
This quantity is two orders of magnitude larger than the current uncertainty in Mars rotation rate estimate and has thus to be removed for any geophysical interpretation of the latter ($\dot\phi_0$ in Eq.~(\ref{Eqphi}) should be replaced by $\dot\phi_{0}^{\mathrm{local}}$).

The 2-body contribution from the transformation between $\tau$ and $t$ to the Mars rotation model in BCRS, including the linear term and the four largest periodic terms, is
\begin{eqnarray}
[\phi]_{\mathrm{2body}} &=& \dot\phi_{0}^{\mathrm{local}} \, [\tau\!-\!t]_\mathrm{GR} 
= \dot\phi_\mathrm{GR} \, t + \sum_j \phi_{rj} \sin j\,l' \, , 
\nonumber \\
&=& 2^\circ.03 \times 10^{-6}\ \da^{-1} \times t \nonumber \\
  && - 166.950\ \mas \sin l' -7.782\ \mas \sin 2l' 
\nonumber \\
&& -0.547\ \mas \sin 3l' - 0.045\ \mas \sin 4l'\, . 
\label{Eq2body}
\end{eqnarray}
These values will be refined in the further subsections considering a more accurate modeling of Mars' trajectory. Nevertheless, the Keplerian modeling presented here is sufficient to get an estimate of the order of magnitude of the impact of the time transformation on Mars rotation.

\subsubsection{Contribution related to the motion of the Sun with respect to the Solar System Barycenter (SSB)}
\label{Section312}

The two-body problem calculation performed in the previous section considers one test mass orbiting one massive body and assume that the coordinate time TDB is the one related to a coordinate system where the massive body is at rest at the origin. For BCRS, this is not the case: the Sun is not at rest and not located at the origin of the coordinate system which is defined as the SSB. Therefore, if one is interested in computing the evolution of Mars’ proper time with respect to TDB, one should use 
\begin{equation}
\frac{\dd\tau}{\dd t}-1=\frac{\LB}{1-\LB}-\frac{1}{1-\LB}\frac{v^2_{M}}{2c^2}-\frac{1}{1-\LB}\frac{GM_\odot}{r_{M\odot}c^2}\, ,
\end{equation}
where $v_M$ is Mars velocity with respect to the SSB and $r_{M\odot}$ is the distance between Mars and the Sun. The only difference with respect to Eq.~(\ref{eq:dtau_dt_2body}) relies in the fact that in Eq.~(\ref{eq:dtau_dt_2body}),
the velocity used is the one of Mars relative to the Sun and not to the SSB. A simple calculation using $\mathbf v_M=\mathbf v_{M\odot} + \mathbf v_{\odot}$ where $\mathbf v_{M\odot}$ is the velocity of Mars with respect to the Sun and $\mathbf v_{\odot}$ is the Sun velocity with respect to the SSB shows that there is an additional contribution to the evolution of Mars' proper time due to the velocity of the Sun with respect to the SSB. This additional contribution is given by
\begin{equation}
 \left[\frac{\dd \tau}{\dd t}\right]_\mathrm{SSB} = - \frac{1}{1-\LB}\frac{\mathbf v_\odot \cdot \mathbf v_{M\odot}}{c^2} -\frac{1}{1-\LB} \frac{v_{\odot}^2}{2c^2} \approx - \frac{\mathbf v_\odot \cdot \mathbf v_{M\odot}}{c^2} \, \cdot
\label{eq:dtau_dt_SSB}
\end{equation}

To first order, the motion of the Sun with respect to the SSB is due to its gravitational interaction with Jupiter. If we take a simple toy model and consider that the motion of the Sun with respect to the SSB is due to Jupiter only and consider Jupiter's orbit to be circular around the Sun, then the Sun's velocity is given by $\vec v_\odot \approx a_J n_J \frac{M_J}{M_\odot}\left(\sin n_J t, -\cos n_J t,0\right)$, where $a_J$ is Jupiter's semi-major axis and $n_J$ its mean motion. To first approximation, we can also consider the orbital motion of Mars to be circular. Eq.~(\ref{eq:dtau_dt_SSB}) can then be integrated analytically
\begin{equation}
\left[\tau\!-\!t\right]_\mathrm{SSB} \approx \frac{aa_J}{c^2}\frac{nn_J}{n-n_J} \frac{M_J}{M_\odot} \sin\big( (n-n_J)\, t + \delta\big)\, ,
\end{equation}
where $a$ is Mars semi-major axis, $n$ its mean motion and $\delta$ the phase difference between the two planets, assuming co-planar motion. The velocity of the Sun with respect to the SSB therefore induces an additional modulation to the $\tau$ to TDB transformation. The period of this modulation is the Mars-Jupiter synodic orbital period, i.e.~$2.235$ years and its amplitude is $37.59$ $\mu$s.

Using Eq.~(\ref{eq:phi_GR}), this modulation will impact the BCRS Mars rotation modeling and induces a modulation of amplitude of $0.55$ mas at the Mars-Jupiter synodic orbital period. A similar calculation considering Saturn leads to a harmonic term of period of $2$ years with an amplitude of $0.11$ mas. The other planets induce periodic terms with amplitudes $\lesssim 0.01$ mas. This can be seen as an indirect effect of the other planets of the Solar system on Mars proper time since it comes from the impact of other planets on the SSB velocity. The direct effect will be computed below.

\subsubsection{Direct contribution from other planets}
\label{Section313}
As can be noticed from Eq.~(\ref{Eqtau2}), the gravitational potential from the other planets will also impact the evolution of Mars' proper time. To first order, the impact from the planets gravitational potential is governed by
\begin{equation}
\left[\frac{\dd \tau}{\dd t}\right]_\mathrm{P} = -\frac{1}{1-\LB}\, \frac{GM_P}{c^2 r_{MP}} \, ,
\label{eq:dtau_planets}
\end{equation}
where $r_{MP}=\Vert\mathbf x_M-\mathbf x_P\Vert$ is the distance between Mars and the planet P.

A simple toy model considering both Mars and the planet to be orbiting on coplanar circular orbits shows that, to first order, the integration of the previous equation leads to a linear drift whose linear coefficients is given by $-GM_P/((a^2+a_P^2)^{1/2} \, c^2)$ and to an harmonic signal at the planet-Mars synodic orbital period and of amplitude of $GM_P\frac{a a_P}{(a^2+a_P^2)^{3/2}}/(c^2(n-n_P))$, where $a_P$ and $n_P$ are the semi-major axis and mean motion of the planet $P$. This calculation is valid only to first order in $a \, a_P/(a^2+a_P^2)$ and neglecting $\LB$. Other harmonics can be identified at higher orders and for non-zero eccentricities, in particular an oscillation at the orbital period of the planet (see Sect.~\ref{sec:VSOP}).

For each planet P, these contributions to the $\tau$ to TDB transformation will impact Mars' BCRS rotation modeling through Eq.~(\ref{eq:phi_GR}), i.e. will produce one term with linear rate $\dot\phi_\mathrm{GR}$ and one harmonic synodic term. For Jupiter, $\dot\phi_\mathrm{GR} =-6^\circ \times 10^{-10}\, \da^{-1}$ (or $-0.00220\ \mas\, \da^{-1}$). For Saturn, $\dot\phi_\mathrm{GR} =-1^\circ \times 10^{-10} \, \da^{-1}$ (or $-0.00037 \ \mas\, \da^{-1}$). The synodic terms associated to Jupiter and Saturn have amplitudes of $0.077$ mas and $0.007$ mas, respectively. The other planets induce synodic terms with amplitudes $\lesssim 0.001$ mas. Though the direct synodic terms associated to Jupiter and Saturn have small amplitudes, they cannot be neglected as they combine with the indirect synodic terms obtained in the previous subsection. The total synodic terms related to Jupiter and Saturn have $0.47$ mas and $0.10$ mas of amplitude, respectively, as the direct and indirect terms are out of phase to each other. The sum of the direct and indirect effect of the planets will be refined in the further subsections considering a more accurate modeling of the planets' trajectory.

Similarly as for the effect of the planets, it is possible to build a toy model for the direct effect of Phobos and Deimos, and of Ceres, the largest body of the asteroid belt. However, given their small mass, the associated relativistic corrections can be neglected.

\subsection{Numerical solution using the DE planetary ephemerides}
\label{Section32}

In this section, we present the result of a numerical integration of Eq.~(\ref{Eqtau2}) using the DE440 planetary ephemerides \citep{park:2021wd} provided by the NAIF-SPICE software \citep{acton:1996fk,acton:2018aa}. The integration is performed starting from J2000 and is performed $30$ years backward and forward. 

In a second step, analytical series consisting in various harmonic terms are fitted to the result of the numerical integration. This procedure is similar to the one developed to produce the Time Ephemeris for Earth, see \citet{fukushima:1995uq,irwin:1999fk,harada:2003vv,fukushima:2010yu}.

\subsubsection{First order contributions}
\label{secDE}

In this section, we will consider the leading contributions from Eq.~(\ref{Eqtau2}), i.e. the $1/c^2$ contributions from the Sun, the various planets and from $\LB$. This integration therefore includes all the effects presented in the previous section. We integrate numerically Eq.~(\ref{Eqtau2}) and transform the evolution of $\tau\!-\!t$ into an estimate of the evolution of $\phi$ through Eq.~(\ref{eq:phi_GR}). We then fit the expression
\begin{equation}
\left[\phi\right]_\mathrm{GR}\!(t) = \dot\phi_\mathrm{GR} \, t + \sum_j (C_j \cos f_j \, t + S_j \sin f_j \, t)
\end{equation}
to the numerically integrated evolution of $\left[\phi\right]_\mathrm{GR}\!(t)$. The values of the various coefficients $\dot\phi_\mathrm{GR}$, $C_j$ and $S_j$ are obtained using a standard linear least-squares fit. Motivated by the toy model presented in the previous section, the angular frequencies $f_j$ included in the fit are chosen as linear combinations of the planets mean motion. We identify the relevant frequencies by (iteratively) searching for the largest peaks in the Fourier transform of the numerically integrated $\left[\phi\right]_\mathrm{GR}\!(t)$ time series. The fitted coefficients are then transform to obtain the following expression
\begin{equation}
\left[\phi\right]_\mathrm{GR}\!(t) = \dot\phi_\mathrm{GR} \, t + \sum_j A_j \sin  (f_j \, t + \varphi_j )\, \cdot
\label{eq:phi_sin}
\end{equation}
The estimated linear term is $\dot\phi_\mathrm{GR}=2^\circ.03021\,\times 10^{-6}\ \da^{-1}$ (or $7.308758\ \mas\ \da^{-1}$) while the amplitude and phase of the harmonic terms are given in Table~\ref{tab:DE}. 
We searched for terms with amplitude down to $0.04$ mas, covering the amplitude range considered in Eq.~(\ref{Eq2body}) for the terms at harmonics of the Martian orbital period. For these terms, the estimated solution is in good agreement with the solution of the toy model (difference $<0.005$ mas), but differs significantly with respect to Eq.~(21) of \citet{YS97}, reminded in Eq.~(\ref{EqYS97}), (up to $9$ mas, or $5\%$ in annual amplitude). The differences are mainly likely due to truncation errors in the parameters values used by \citet{YS97}. The estimated terms at the Mars-Jupiter and Mars-Saturn synodic periods are $0.10$ mas and $0.005$ mas larger than obtained with the toy model, as a result of the assumption of circular planetary orbits therein. We also find one term at the orbital period of Jupiter with an amplitude of about $0.04$ mas, related to the direct effect of the planet, and 3 others terms at different periods with amplitude ranging between $0.06$ mas and $0.08$ mas, mainly due to the indirect effect of the Earth and of Jupiter of the orbit of Mars. \\

\begin{figure}[!htb]
\centering
\includegraphics[width=0.5\textwidth]{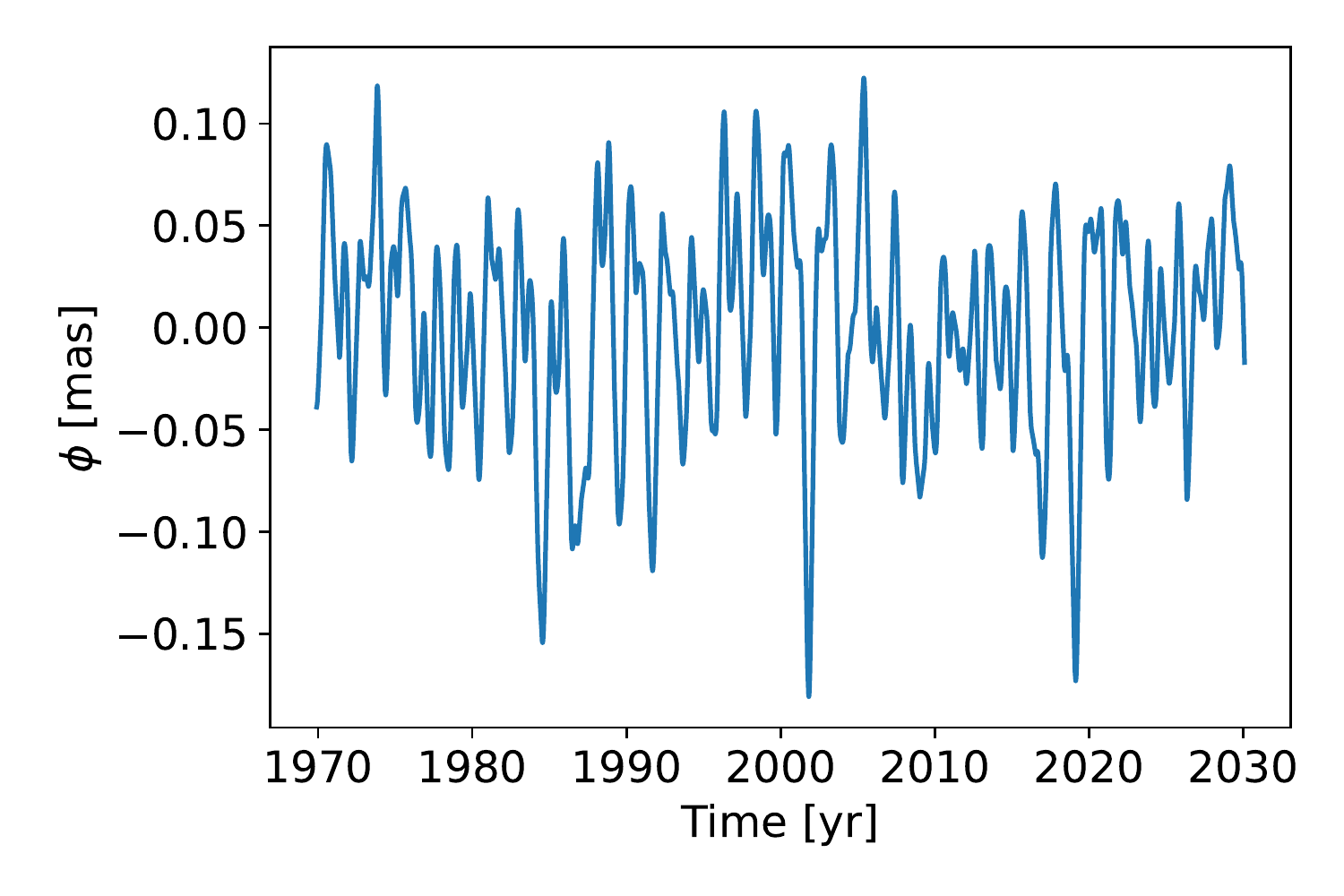}
\caption{Difference between the $\left[\phi\right]_\mathrm{GR}\!(t)$ obtained by integrating numerically Eq.~(\ref{Eqtau2}) using the DE440 planetary ephemerides and using Eq.~(\ref{eq:phi_GR}) and the fitted series from Eq.~(\ref{eq:phi_sin}). This curve provides an estimate of the accuracy of the fitted analytical model provided by the coefficients from Table~\ref{tab:DE}.}\label{fig:res_DE}
\end{figure}

The residuals between the numerical integration and the fitted harmonics decomposition is presented in Fig.~\ref{fig:res_DE} and remains below $0.15$ mas. Formal uncertainties are not relevant quantities to characterise the errors of the fit as they do not directly rely on any observations (no data points). Instead, we use another method described in Sec. \ref{sec:comparison} to compute the accuracy of the estimated coefficients which equals $\sim 0.01$ mas on the main terms.

\begin{table*}
 \caption{Coefficients parametrizing the evolution of $\left[\phi\right]_\mathrm{GR}\!(t)$ using Eq.~(\ref{eq:phi_sin}) fitted on the numerical integration of Eq.~(\ref{Eqtau2}) using the DE440 planetary ephemeris and using Eq.~(\ref{eq:phi_GR}).} 
\label{tab:DE}
\centering
\begin{tabular}{cccc}
\hline \hline
$2\pi/f_j$ (yr) & Amp (mas) & Phase ($^\circ$) & Comment \\
\hline
 0.470223 & 0.045 & 257.459 & 1/4 Mars orbital period \\
 0.626964 & 0.544 & 238.143 & 1/3 Mars orbital period \\
 0.940446 & 7.783 & 218.770 & 1/2 Mars orbital period \\
 1.11764 & 0.077 & 101.380 & 1/2 Mars-Jupiter synodic period \\
 1.880892 & 166.949 & 199.384 & Mars orbital period \\
 2.00913 & 0.097 & 302.306 & Mars-Saturn synodic period \\
 2.23528 & 0.567 & 321.360 & Mars-Jupiter synodic period \\
 2.7543 & 0.075 & 63.492  & Jupiter-Mars 2-1 resonance \\ 
 11.862 & 0.038 & 157.048 & Jupiter orbital period \\
 15.781 & 0.060 & 292.431 & Mars-Earth 2-1 resonance \\
\hline
\end{tabular}
\end{table*}

\subsubsection{Higher order contributions}

\begin{figure}[!htb]
\centering
\includegraphics[width=0.5\textwidth]{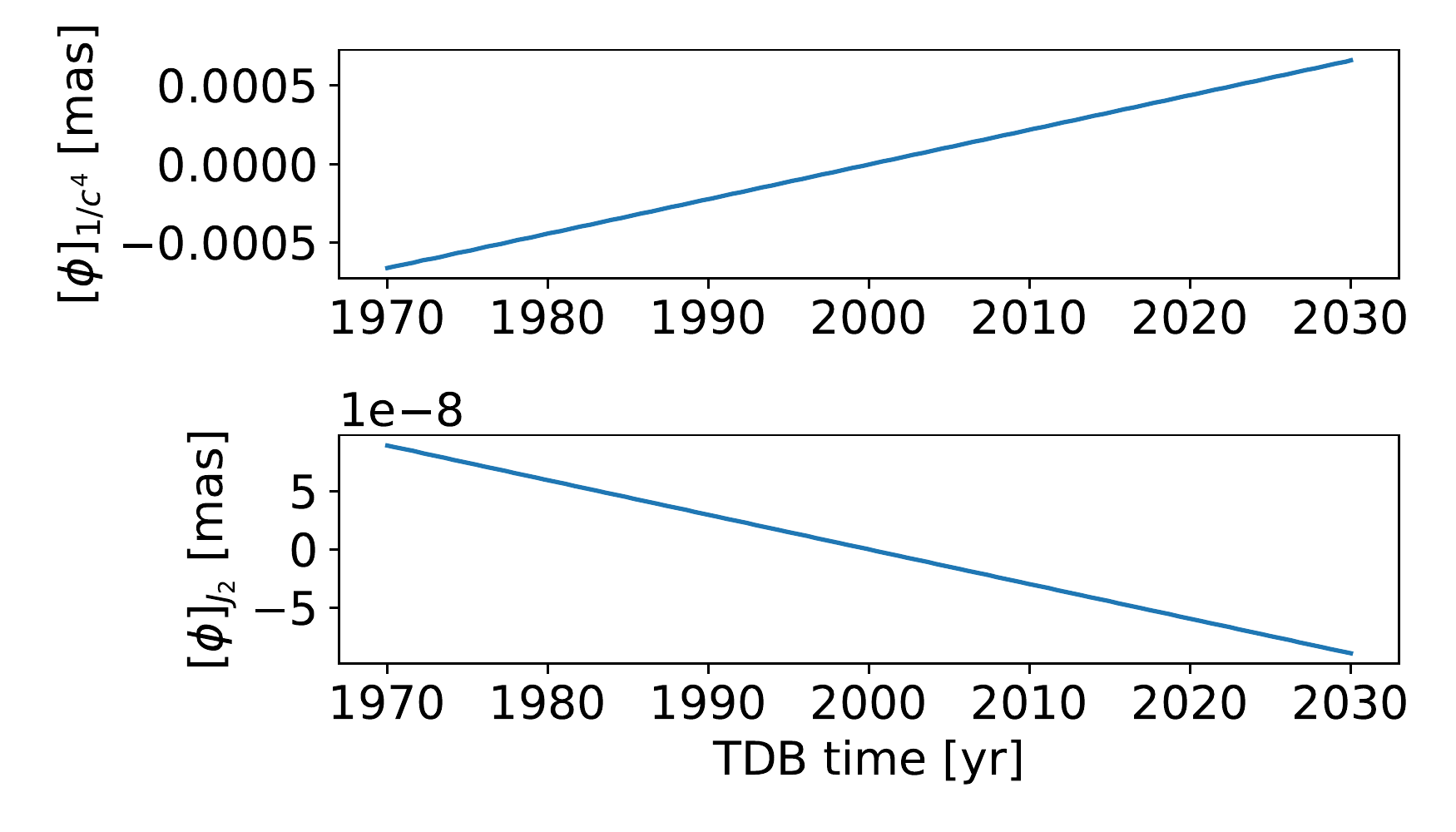}
\caption{Top: impact of the $1/c^4$ term from Eq.~(\ref{Eqtau2}) on the BCRS modeling of Mars' rotation. 
  Bottom: impact of the Sun quadrupole moment ($J_2^\odot$) on the BCRS modeling of Mars' rotation.}\label{fig:higher_order}
\end{figure}

In this section, we consider the $1/c^4$ contribution appearing in Eq.~(\ref{Eqtau2}) and the contribution from the Sun's quadrupole moment $J_2^\odot$. Their impact on the BCRS modeling of Mars' rotation is presented in Fig.~\ref{fig:higher_order} and is of the order of $\mu$as for the $1/c^4$ term and of the order of 10 nas for the $J_2^\odot$. Both these contributions can safely be neglected.

\subsection{Series solution using VSOP ephemerides}
\label{sec:VSOP}

In this section, we present the results of a semi-analytical approach based on series for the barycentric position and distance of planets as provided by the analytical planetary theory VSOP87 \citep{VSOP87}. We integrate Eq.~(\ref{Eqtau2}), neglecting the $1/c^4$ and Sun's $J_2^\odot$ contributions, identify the harmonics with the largest contribution to the $\tau\!-\!t$ relationship, and estimate the evolution of $\phi$ through Eq.~(\ref{eq:phi_GR}). In a second step, we numerically assess the accuracy of the semi-analytical solution.\\

\subsubsection{Semi-analytical solution}

In VSOP87, the barycentric Cartesian coordinates $(X,Y,Z)$ and distance $r$ of the planets to the Sun are written as series of the form 
\begin{subequations}
\begin{eqnarray}
&& \sum_j\left( C_{j} \cos\varphi_{j} + S_{j} \sin\varphi_{j}\right)\, ,
\label{SeriesForm} \\
&& C_j=\sum_{\alpha} T^\alpha C_{\alpha,j}\, , \\
&& S_j=\sum_{\alpha} T^\alpha S_{\alpha,j}\, ,
\end{eqnarray}
\end{subequations}
where $C_{\alpha,j}$ and $S_{\alpha,j}$ are amplitudes and $T$ is the time measured in thousands of Julian years from J2000. $\varphi_{j}$ are linear combinations of fundamental arguments, including the mean longitudes of Saturn ($S\!a$), Jupiter ($J\!u$), Mars ($M\!a$), and the Earth ($T\!e$), see Table~2 of \citep{VSOP87}:
\begin{subequations}
\begin{eqnarray}
S\!a&=&0.87401675650 + 213.2990954380\,T\, , 
\label{Sa} \\
J\!u&=&0.59954649739 + 529.6909650946\,T\, , \\
M\!a&=&6.20347611291 + 3340.6124266998\,T\, ,
\label{Ma} \\
T\!e&=&1.75347045953 + 6283.0758499914\,T\, \cdot
\end{eqnarray}
\end{subequations}
The power $\alpha$ is an integer in-between 0 and 5. For $\alpha=0$, the series are periodic. For $\alpha\geq1$, the series are pseudo-periodic (Poisson series). 

The solution for $[\phi]_\mathrm{GR}(t)$ is firstly written as
\begin{subequations}
\begin{eqnarray}
[\phi]_\mathrm{GR}(t) &=&\dot\phi_\mathrm{GR} \times t + \sum_j \left(\Delta\phi_j^{c} \cos\varphi_j + \Delta\phi_j^{s} \sin\varphi_j \right)\, , 
\label{SolForm} \\
\Delta\phi_j^{c/s} &=&\sum_{\alpha} T^\alpha \Delta\phi_{\alpha,j}^{c/s}\, ,
\end{eqnarray}
\end{subequations}
with $\Delta\phi_j^{c/s}$ the amplitudes of the periodic and Poisson series. In the first place, the fundamental arguments of the series will be the same as the VSOP87 arguments, by construction, because
the series for $v^2$, the squared Mars' barycentric velocity, is directly obtained as the squared norm of the time derivative of the position vector of Mars $(X,Y,Z)$, while the series for $1/r_{\odot}$ is obtained starting with the VSOP series for $r_{\odot}$ and following Eq.~(61) of \citet{BMAN20}. The distance $r_{P}$ between Mars and another planet varies greatly with time, and as a result, it is difficult to express as convergent series for $1/r_{P}$ starting from the VSOP series for the Cartesian coordinates. We therefore assume that the orbits of Mars and of the other planets are Keplerian and coplanar, and use the mean orbital elements of \citet{simon:2013vz}. We adapt the procedure described in Sect.~4.3.3 of \citet{BMAN20} to obtain a series for $1/r_P^5$ to the case of $1/r_P$. This can be seen as extension of the toy model presented in Sect.~\ref{Section313} to higher orders in eccentricities and in $a \, a_p/(a^2+a_p^2)$ and, as a result, we will identify more harmonics and obtain different amplitudes.

Then, for consistency with the usual form of $[\phi]_\mathrm{GR}(t)$, expressed with the mean anomaly $l'$ of Mars as the argument (see Eq.~\ref{EqYS97}), we change the fundamental arguments, using the mean anomalies of the planets
\begin{subequations}
\label{ll}
\begin{align}
&l_{S\!a} = S\!a-\varpi_{S\!a}\,=\,5.53304687684+213.2002152909 \, T\, , 
\label{lSa} \\
&l_{J\!u} = J\!u-\varpi_{J\!u}\,=\,0.52395267692+529.6533496052 \, T\, ,\\
&l'=M\!a-\varpi_{M\!a}\,=\,0.3381185455+3340.5349512479 \, T\, , 
\label{lMa} \\
&l_{T\!e}=T\!e-\varpi_{T\!e}\,=\,6.24006011944+6283.0195517158\,T\, , \label{lTe} 
\end{align}
\end{subequations}
instead of their mean longitudes. We express $[\phi]_\mathrm{GR}(t)$ correct up to the first order in the rates of the pericenter longitudes $\varpi$ of the planets, creating a second Poisson series, to add to the first Poisson series coming directly from VSOP ephemerides, and which is not affected by the argument change, at first order. Both Poisson series are similar, but with opposite amplitudes, and therefore almost cancel each other (the sum of the two series is smaller than $0.05$ mas on the interval $\pm 30$ years around J2000, and significant Poisson terms were not found in the fit of the numerical solution). As a result, we omit Poisson series in the following. 

Finally, the periodic series in $[\phi]_\mathrm{GR}(t)$ is written in a pure Sine form, convenient for application purpose:
\begin{equation}
[\phi]_\mathrm{GR}(t) = \dot\phi_\mathrm{GR}\, t + \sum_j \phi_{j}^r \sin(f_j \, t + \varphi_{j}^0)\, ,
\label{SolFormSine}
\end{equation}
with $\phi_{j}^r$ the amplitudes and $\varphi_j^0$ a phase (different from the one of Eq.~\ref{SolForm}). 
$f_j$ are linear combination of the rate of the mean anomalies of Eqs.~(\ref{ll}).

For the constant rate, we find $\dot\phi_\mathrm{GR} = 7.3088\ \mas \ \da^{-1}$, in agreement with the sum of the respective contributions from $L_B$, from the Sun, and from each planet as obtained with the fit of the numerical solution. The periodic terms, with amplitudes down to $0.04$ mas, are presented in Table~\ref{tab:VSOP}. 
We find the same 10 periodic terms as with the fit of the numerical solution of Table~\ref{tab:DE}, but also one additional term, at the orbital period of Saturn ($\sim30$ years) and with an amplitude of $0.043$ mas. We did not find this term with the fit because of its long period and of its small amplitude.

\begin{table*}
\caption{Terms of the periodic series of Eq.~(\ref{SolFormSine}), down to $0.04$ mas in amplitude using the VSOP ephemerides.
For each term, the frequency $f_j$ is obtained as the rate of the linear combination of the mean anomalies as described in Columns 2-5.} 
\label{tab:VSOP}
\centering
\begin{tabular}{cccccccc}
\hline \hline
$j$ & $l_{S\!a}$ & $l_{J\!u}$ & $l'$ & $l_{T\!e}$ & $2\pi/f_j$(y) & Amp (mas) & Phase ($^\circ$) \\
\hline
1 & 0 & 0 & 4. & 0 & 0.470223 & 0.045 & 257.492 \\
2 & 0 & 0 & 3. & 0 & 0.626964 & 0.544 & 238.119 \\
3 & 0 & 0 & 2. & 0 & 0.940446 & 7.783 & 218.746 \\
4 & 0 & -2. & 2. & 0 & 1.117654 & 0.077 & 101.316 \\
5 & 0 & 0 & 1 & 0 & 1.880892  & 166.958 & 199.373 \\
6 (Syn) & -1. & 0 & 1 & 0 & 2.009118  & 0.107 & 305.198 \\
7 (Syn) & 0 & -1 & 1. & 0 & 2.235308 & 0.566 & 320.635 \\
8 & 0 & -2. & 1. & 0 & 2.754299 & 0.075 & 66.885 \\
9 & 0 & 1 & 0 & 0 & 11.862826 & 0.043 &  148.17 \\
10 & 0 & 0 & 2. & -1 & 15.784901 & 0.060 & 290.792 \\
11 & 1 & 0 & 0 & 0 & 29.470821 & 0.043 & 120.366 \\
\hline
\end{tabular}
\end{table*}

\subsubsection{Accuracy of the semi-analytical solution}
\label{sec:comparison}

\begin{figure*}[!htb]
\centering
\includegraphics[width=1.\textwidth]{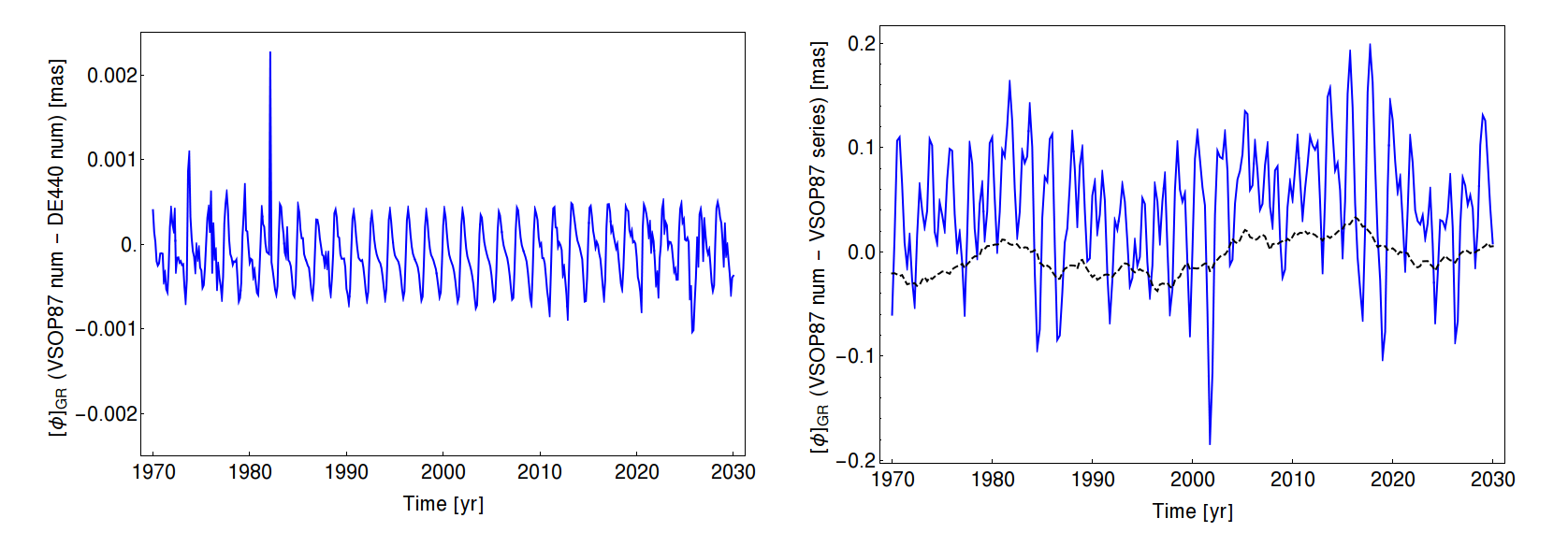}
\caption{Left: Difference between the solutions for $[\phi_\mathrm{GR}](t)$ obtained using VSOP87 and DE440 ephemerides (numerical integrations). 
Right: Difference between the numerical integration and the semi-analytical solution using VSOP87 (with only the 11 terms of Table~\ref{tab:VSOP} in blue, with all terms in black).}\label{Fig_diff}
\end{figure*}

The difference between the numerical integrations of $[\phi_\mathrm{GR}](t)$ performed using the recent DE440 and the older VSOP87 ephemerides is presented in the left panel of Fig.~\ref{Fig_diff} and remains below $0.003$~mas, indicating that using the VSOP87 theory should not be a cause of major errors.

The different steps of the computational procedure to obtain the semi-analytical solution of Eq.~(\ref{SolForm}) introduce residuals smaller than $0.05$ mas when all the terms of the periodic series are considered, and smaller than $0.2$ mas when only the $11$ largest terms of the series are considered (see right panel of Fig.~\ref{Fig_diff}). 
This is in agreement with the residuals of the fitted solution to the numerical integration based DE440 ephemerides, that includes the 10 periodic terms of Table~\ref{tab:DE}, see Fig. \ref{fig:res_DE}. 

\begin{figure}[!htb]
\centering
\includegraphics[width=0.5\textwidth]{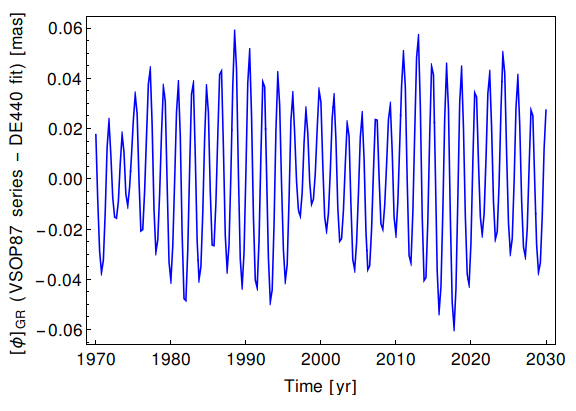}
\caption{Difference between the $[\phi_\mathrm{GR}](t)$ of the semi-analytical solution using VSOP87 (with only the 11 terms of Table~\ref{tab:VSOP}) and the fit of the numerical integration with DE440 ephemerides.} \label{Fig5}
\end{figure}

The amplitudes and phases of the fitted (Table~\ref{tab:DE}) and semi-analytical (Table~\ref{tab:VSOP}) solutions are in good agreement to each other (see also Fig.~\ref{Fig5}, where the difference remains below $0.06$ mas). The differences give a sense of the modeling uncertainties: e.g.~$0.01$ mas ($0.005\%$) in annual amplitude, or $0.01$ mas ($0.9\%$) on the Mars-Jupiter synodic term. Both solutions or an averaged solution can be used for application purpose.

\section{Signatures in the Doppler observable}

As shown in the previous sections, the relativistic variations in Mars rotation and orientation mainly affect the angles $\psi$ and $\phi$. 
Using analytic expressions, we characterise in this section the signature of these variations in the Doppler observable of a Martian lander communicating directly with the Earth. 

Numerical applications are provided for the specific case of RISE (Rotation and Interior Structure Experiment), the radio-science experiment of the NASA InSight mission \citep{Fo18}. The level of the signatures and their temporal behaviour are compared to the noise level and non-relativistic signatures, respectively.

Let us note $\delta(\dot\psi_0 t)$ a variation in the precession,  $\delta(\psi_{nut})$ a variation in the nutation in longitude, and $\delta\phi$ a variation (linear and/or periodic) in the rotation angle $\phi$, excluding the nutation term $\psi_{nut}\cos{\varepsilon_0}$ (see Eqs.~\ref{Eqpsi} and \ref{Eqphi}).  These three notations can be used to represent any kind of variations in $\psi$ and $\phi$, including the relativistic ones. Considering that $\delta(\dot\psi_0 t)$, $\delta(\psi_{nut})$, and $\delta\phi$ are small quantities,
their signature in the observable can be written to the first order as \citep{Yse17}:
\begin{subequations}~
\label{eq_analyti}
\begin{align}
\Delta q_{\delta(\dot\psi_0 t)} &= \delta(\dot\psi_0 t) \, \Omega \, R \, \cos{\theta} \,\nonumber \\
& \bigg(\cos\delta_E \cos H_E \cos\varepsilon - \sin\delta_E \sin\left(H_E+\alpha_E\right) \sin\varepsilon \bigg)\, ,
\label{eq_UTsigna} \\
\Delta q_{\delta(\psi_{nut})} &= -\delta(\psi_{nut}) \,\Omega \, R \, \cos{\theta}\, \sin\delta_E \sin(H_E+\alpha_E)\sin\varepsilon \, ,
\label{eq_UTsignb} \\
\Delta q_{\delta\phi} &=\delta\phi\, \Omega \, R \, \cos\theta\, \cos H_E \, \cos\delta_E \, ,
\label{eq_UTsignc}
\end{align}
\end{subequations}
with $\theta$ the lander latitude,
$R$ the radius of Mars,
$\Omega$ the rotation rate and
$\delta_E$ the Earth declination relative to Mars equator.
Each of these Doppler variations has a diurnal modulation via $H_E$ the Earth hour angle seen from Mars. 
The observable variations defined in Eqs.~(\ref{eq_analyti}) are Doppler shift expressed as the variation of the velocity along the line-of-sight (LOS). For a round-trip (two-way) radio link, the conversion factor between the LOS velocity and the Doppler observable is $2 f_t/c$ with $f_t$ the carrier downlink frequency. In the case of RISE (but also of LaRa, a radio transponder ready to fly to Mars, see \citealt{LaRaPaper}), $f_t\simeq 8.4$ GHz, which is in the X frequency band. 

The signature of $[\phi]_{GR}(t)$ (linear and/or periodic terms) is computed using Eq.~(\ref{eq_UTsignc}). The signature of the geodetic precession in longitude is obtained using Eq.~(\ref{eq_UTsigna}), and that of the geodetic nutation in longitude is computed with Eq.~(\ref{eq_UTsignb}). Precession and nutation signatures differ from each other because $\psi_{nut}$ also affects the angle $\phi$ while $\dot\psi_0$ does not (see Eqs.~(\ref{Eqpsi}) and (\ref{Eqphi})).
 
\begin{figure*}[!htb]
\centering
\includegraphics[width=0.8\textwidth]{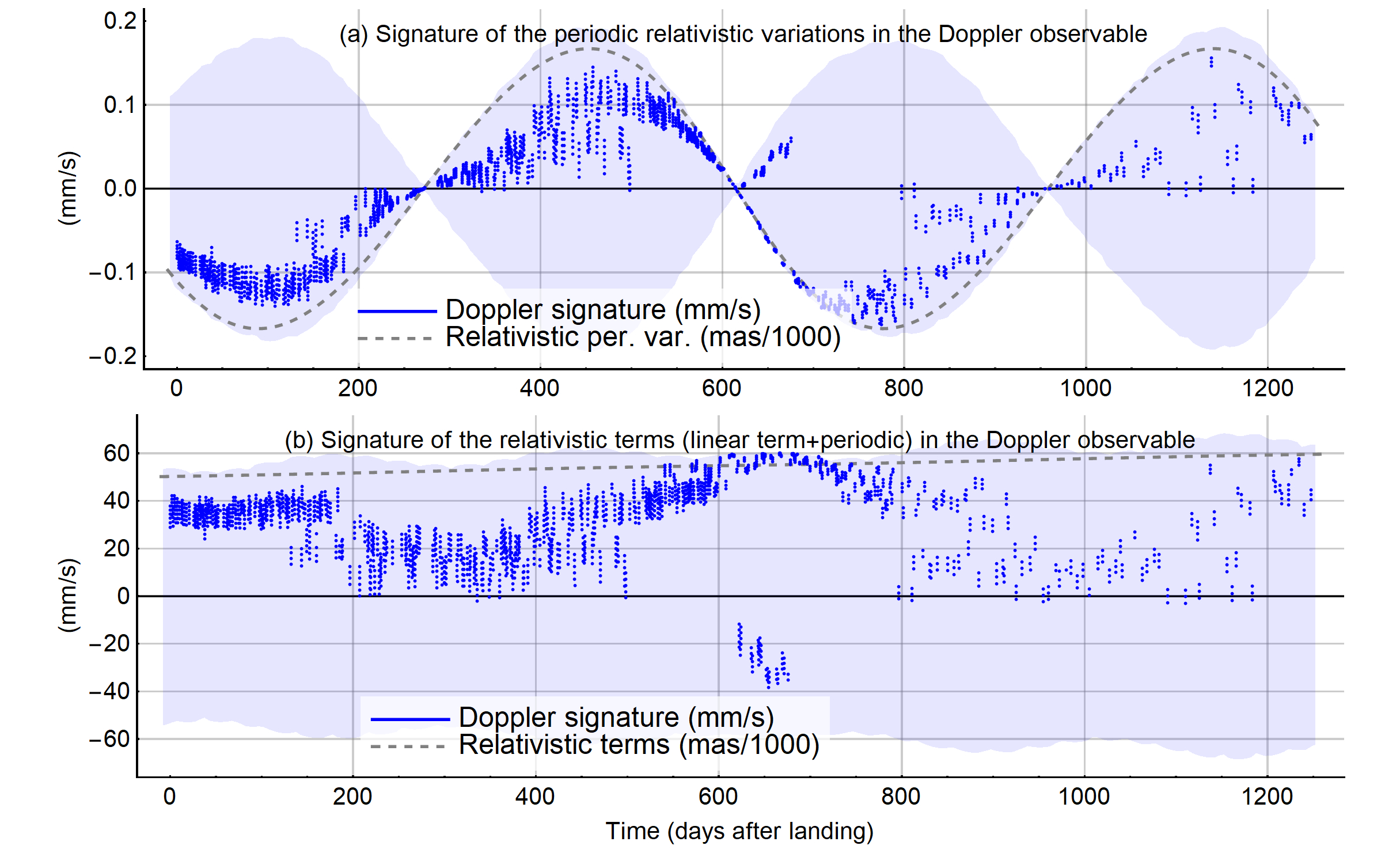} 
\caption{(a) Signature in the Doppler observable (in mm/s) of the periodic relativistic variations ($[\phi]_{GR}(t) - \dot\phi_{GR} \, t$) in the rotation angle as a function of time, using the RISE timing (Nov. 2018-April 2022).
The blue envelope uses a simulated continuous timing. The gray dashed line represents the periodic relativistic variations in the rotation angle $\delta\phi$, arbitrarily rescaled. (b) Signature in the Doppler observable (in mm/s) of both the linear and periodic relativistic terms $[\phi]_{GR}(t)$ in the rotation angle, using the RISE timing. }
\label{figsign}
\end{figure*}

The periodic variations in $\phi(t)$ induced by the seasonal atmosphere/surface dynamics (see for example~\citealt{Kono20})
and by the time coordinates transformation together result in a maximum angular displacement of the lander of $\sim 670$ mas as seen from the center of Mars ($1$ mas corresponds to a displacement of $1.6$ cm at the surface of Mars). 
For a lander located at the InSight landing site (i.e.~Elysium Planitia: $4.5^\circ$N, $135.62^\circ$E, $-2.6$ km altitude), such an angular displacement induces a Doppler shift in the RISE measurements of $\lesssim$
$0.56$ mm/s (see Fig.~2 of \citet{Yse17} or Table~\ref{tabsign}). A fourth of this Doppler signal, computed at the RISE tracking data-timing using the RISE frequency, comes from the relativistic periodic variations (see Fig.~\ref{figsign}a and extra information in Table~\ref{tabsign}). 

\begin{figure*}[!htb]
\centering
\includegraphics[width=0.8\textwidth]{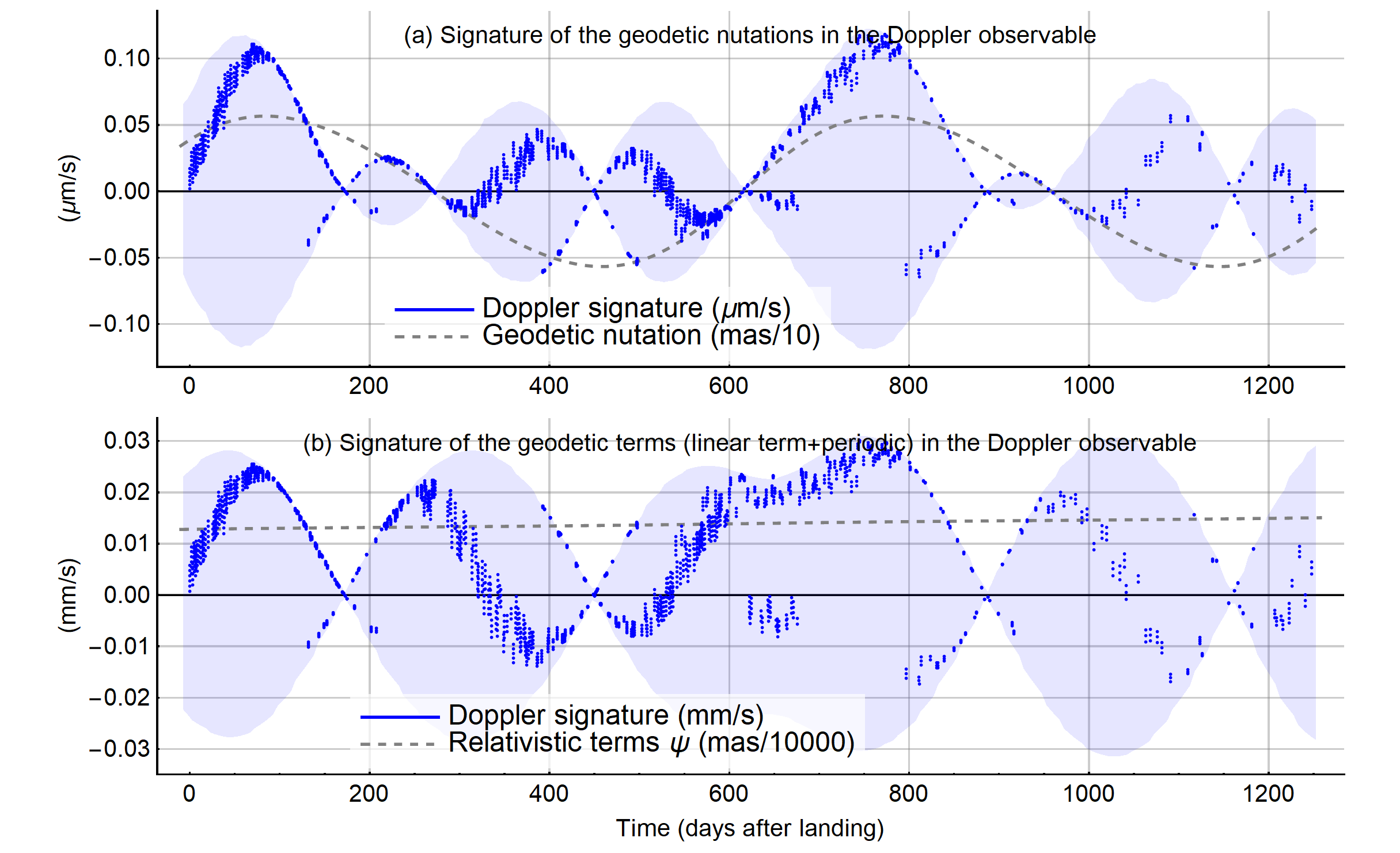} 
\caption{(a) Signature in the Doppler observable (in $\mu$m/s) of the small geodetic nutations in the longitude angle as a function of time, using the RISE timing (Nov. 2018-April 2022). The blue envelope uses a simulated continuous timing. The gray dashed line represents the geodetic nutations, arbitrarily rescaled. 
(b) Signature in the Doppler observable (in mm/s) of all the geodetic terms (nutations and the linear term), using the RISE timing. }
\label{figsign2}
\end{figure*}

The combination of seasonal variations in Euler angles and of diurnal trend in the hour angle produces symmetrical envelopes with respect to zero in the Doppler signature as seen in Figs.~\ref{figsign}~and~\ref{figsign2}. Because of the repeatability in the RISE observation timing imposed by its fixed and directional antennas, the data points cover a limited part of these diurnal cycles. 

\begin{table*}
\caption{Maximal amplitude of the signature in the range and the Doppler observables of the relativistic variations in the rotation angles for a lander in Elysium Planitia and using the RISE timing.}
\label{tabsign}
\centering
\begin{tabular}{ccccc}
\hline \hline
Contribution & Magnitude & Signature in & \multicolumn{2}{c}{Signature in Doppler obs.} \\
& (mas) & range obs. (m)  & (mm/s) &  (mHz) \\
 \hline
Periodic variations in $\phi$ (seasonal and relativistic) & 670 & 10 & 0.56 & 31 \\
$[\phi]_\mathrm{GR}(t)$ (periodic terms only) & 167 & 2.7 & 0.14 & 7.8 \\
$[\phi]_\mathrm{GR}(t)$ (linear and periodic terms) & 58 000 & 950 & 49 & 2800 \\ 
$[\phi]_\mathrm{GR, \, \textrm{this paper}}$ - $[\phi]_\mathrm{GR, \, \textrm{Yoder \& Standish (1997)}}$ 
& 8.8 & 0.14 & 0.007 & 0.41 \\ 
&&& \\
Liquid core contribution to nutations & $\sim$~20-30 & $\sim$~0.3 &$\sim$~0.01 & $\sim$~0.5 \\
$\psi_{so}$ (geodetic nutation only) & 0.6 & 0.001 &  0.0001 & 0.007 \\
$\psi_{so}$ (geodetic nutation and linear term) & 150 & 0.44 & 0.03 & 1.7 \\
&&& \\
InSight/RISE noise level & & & 0.02 & 1.1 \\
\hline
\end{tabular}
\end{table*}

The $\sim10$ mas difference between the periodic terms of our solution for $[\phi]_{GR}$ and that first estimated by \citet{YS97} (see Eq.~\ref{EqYS97} and Sect.~\ref{secDE}) has a Doppler signature lower than $0.007$ mm/s, 
which is smaller than the RISE noise level ($1.1$ mHz at $60$s integration time, corresponding to $0.02$ mm/s) and smaller than the liquid core signature ($\sim~0.01$ mm/s, 
\cite{Yse17, RisePaper}). 
However a precise solution of Mars rotation angles (at the level of $1$ mas or smaller) is needed to correctly interpret the measured periodic variations in term of atmospheric constraints.

The signature in the Doppler of the relativistic linear term of $7.3\ \mas$/day (see Eq.~\ref{eqfinale}) is very large (up to $49$ mm/s) as shown in Fig.~\ref{figsign}b. This large signal with a linear increase, barely visible in the plot, happens because it linearly depends on the time of the observations (2018-2022 for InSight) relative to the chosen reference epoch (here J2000). 

The signature of the geodetic nutation in longitude in the Doppler observable is shown in Fig.~\ref{figsign2}a to be very small ($\le0.12 \, \mu$m/s), while that of the geodetic precession (linear term in Eq.~\ref{EQ13}) is two to three orders of magnitude larger (up to $0.03$ mm/s) as shown in Fig.~\ref{figsign2}b, where the signature of the linear plus periodic geodetic terms is plotted (see Eq.~\eqref{EQ13}). Similarly as for the relativistic linear term in the rotation angle, the larger signature of the geodetic linear term in longitude for Mars ($6.75$ mas/yr) results from the linear dependency to the time past from the chosen origin of time (i.e. $\sim150$ mas in 2020 for a reference epoch at J2000).

\section{Application to other planets}
\label{Section5}

Mars has been extensively explored, but the rotation of other bodies of the Solar system is also subject to investigations. We here provide an estimate of the main terms of the relativistic correction in longitude $\psi$ and rotation $\phi$ to include in the rotation model of our neighboring planets. For each planet, we first estimate the geodetic precession and nutation in longitude. Then we estimate the linear and periodic changes in rotation, considering (1) the Sun contribution for a planet on a Keplerian orbit, (2) the contribution related to the Sun motion with respect to the SSB, and (3) the direct contribution from the other planets.

\subsection{Geodetic precession and nutations}
For any planet, the spin-orbit angular velocity $\bm{\Omega}_{\mathrm{so}}$ of Eq.~(\ref{eq:Omega_SO}) is proportional to $\textbf{x} \times \textbf{v}$ and therefore perpendicular to the orbital plane (see also Eq.~\eqref{EqOmegaso}). As a result, if a planet moves on a Keplerian orbit, only its longitude angle (defined as the longitude of the spin axis with respect to the orbital plane) is affected by the geodetic precession and nutation, whereas the obliquity and rotation angles $\varepsilon$ and $\phi$ are unchanged:
\begin{subequations}
\begin{eqnarray}
\dot\psi_{\mathrm{so}} &=& \Omega_{\mathrm{so}}^z \, ,
\label{EQ43a}\\
\dot\varepsilon_{\mathrm{so}} &=& 0\, ,\\
\dot\phi_{\mathrm{so}} &=& 0\, .
\end{eqnarray}
\end{subequations}
The geodetic precession and nutation for each planet, as obtained with the toy model of Sect.~\ref{SectRelPRec}, is given in Table~\ref{Tab5}. The geodetic precession rate increases with decreasing distance to the Sun. The amplitudes of the periodic terms do not strictly follow that rule of thumb because they are relatively more dependent on eccentricity.\\

\begin{table*}[!ht]
 \caption{Geodetic precession and nutation in longitude for the planets of the Solar System (in mas, $t$ is the time in years).}
\label{Tab5}
\centering
\begin{tabular}{cl}
\hline \hline
Planet & $\psi_{\mathrm{so}} (t)$\\
\hline
Mercury  & $214.887\ \times t + 4.996 \sin{l_{M\!e}} + 0.759\ \sin{2l_{M\!e}} + 0.151\ \sin{3l_{M\!e}}$ \\
Venus    & $\phantom{0}43.123 \times t + 0.086\ \sin{l_{V\!e}}$ \\
Earth    & $\phantom{0}19.193 \times t + 0.153\ \sin{l_{T\!e}} + 0.002\ \sin{2l_{T\!e}}$ \\
Mars     & $\phantom{00}6.754 \times t + 0.565\ \sin{l'}\, \, \,+ 0.039\ \sin{2l'}   \quad + 0.004\ \sin{3l'}$ \\
Jupiter  & $\phantom{00}0.312 \times t + 0.086\ \sin{l_{J\!u}} + 0.003\ \sin{2l_{J\!u}}$ \\
Saturn   & $\phantom{00}0.068 \times t + 0.053\ \sin{l_{S\!a}} + 0.002\ \sin{2l_{S\!a}}$ \\
Uranus   & $\phantom{00}0.012 \times t + 0.022\ \sin{l_{U\!r}} + 0.001\ \sin{2l_{U\!r}}$ \\
Neptune  & $\phantom{00}0.004 \times t + 0.003\ \sin{l_{N\!e}}$ \\
\hline
\end{tabular}
\end{table*}

For the Earth, we obtain consistent results with \citet{fukushima91}, as we follow the same approach based on a Keplerian orbit (see also Eq.~(27) of \citealp{soffel:2003bd}). Even though their solution is presented as an approximation, our results for the precession and annual terms of the eight planets are in a very good agreement with those presented in Table~1 of \citet{Eroshkin2007} who fitted a solution to a numerical integration based on ephemerides. This is because of two approximations which compensate each other during their computation: (1) they refer the geodetic motion of all planets to the Earth ecliptic of J2000, instead of their respective orbital plane, 
(2) they neglect the equatorial components of the angular velocity vector $\vec \sigma = (\sigma^X, \sigma^Y,\sigma^Z)$ expressed in the coordinates of a frame attached to the Earth ecliptic of J2000. For the demonstration, we first write the geodetic variations in ``Ecliptic Euler angles'' (we use the notation $\ast$ for these angles) as
\begin{subequations}
\begin{eqnarray}
\dot \psi^{\ast}_{\mathrm{so}} &=& \sigma^Z-\sigma^X \sin\psi^{\ast}_0 \cot \varepsilon^{\ast}_0+\sigma^Y \cos\psi^{\ast}_0 \cot \varepsilon^{\ast}_0\, ,
\label{EQ44a} \\
\dot\varepsilon^{\ast}_{\mathrm{so}} &=& \sigma^X \cos\psi^{\ast}_0+\sigma^Y \sin\psi^{\ast}_0\, ,\\
\dot\phi^{\ast}_{\mathrm{so}} &=& (\sigma^X \sin\psi^{\ast}_0-\sigma^Y \cos\psi^{\ast}_0) / \sin\varepsilon^{\ast}_0\, \cdot
\end{eqnarray}
\end{subequations}
Then we write $\vec\sigma=R_z(-\Omega_0)\cdot R_x(-i_0)\cdot\bm{\Omega}_{\mathrm{so}}$, with $\Omega_0$ and $i_0$ the ascending node longitude and inclination of the planet's orbit with respect to the ecliptic. Since $\Omega_{\mathrm{so}}^x\simeq0$ and $\Omega_{\mathrm{so}}^y\simeq0$ (for a non Keplerian orbit, small variations about zero are possible), $\sigma^X\simeq \Omega_{so}^z \sin i_0 \sin\Omega_0$, $\sigma^Y\simeq - \Omega_{\mathrm{so}}^z \sin i_0 \cos\Omega_0$, and $\sigma^Z\simeq\Omega_{\mathrm{so}}^z \cos i_0$. To obtain $\psi_{\mathrm{so}}(t)$, we integrate $\Omega_{\mathrm{so}}^z$ over time (see Eq.~\ref{EQ43a}). For bodies with small orbital inclination with respect to the Earth ecliptic, $\sigma^Z\simeq \Omega_{\mathrm{so}}^z$ and by integrating only $\sigma^Z$ while neglecting $\sigma^X$ and $\sigma^Y$ in Eq.~(\ref{EQ44a}), \citet{Eroshkin2007} obtain in fact $\psi_{\mathrm{so}}(t)$ instead of $\psi^\ast_{\mathrm{so}}(t)$. 

In subsequent studies (e.g.~\citealp{pashkevich2016}), the geodetic variations in Ecliptic Euler angles, taking into account $\sigma^X$ and $\sigma^Y$ were described. We here compute the precession rate in ecliptic Euler angles for Mars, by multiplying $6.754$ mas/y by $\left(\cos i_0 - \sin i_0 \cot \varepsilon^\ast_0 \cos(\psi^\ast_0-\Omega_0)\right)$, $- \sin i_0 \sin(\psi^\ast_0-\Omega_0)$, and $(\sin i_0 \cos(\psi^\ast_0-\Omega_0))/\sin\varepsilon^\ast_0$, respectively. With $i_0=1^\circ.84973, \Omega_0=49^\circ.5581, \psi^\ast_0=82^\circ.9071, \varepsilon^\ast_0=26^\circ.7179$, we find $6.389$, $-0.120$ and $0.405\ \mas\ \yr^{-1}$ in $\psi^\ast, \varepsilon^\ast$ and $\phi^\ast$, respectively. For the rate in obliquity and rotation angle, we obtain values consistent with \citet{pashkevich2016}, but with opposite sign. The geodetic rate in longitude of \citet{pashkevich2016} is $7.114\ \mas\ \yr^{-1}$. We believe this value was obtained by mistake because of a sign confusion for the ecliptic obliquity $\varepsilon^\ast_0$. 

\subsection{Rotation variations due to time coordinate transformation}
\label{Section5.2}

In Table~\ref{Tab6}, we present estimates for the Sun contribution to the relativistic variations in $\phi$, for a rotation model expressed in the BCRS, assuming that the planets follow Keplerian orbits (see toy model of Section \ref{Section311}). This 2-body contribution tends to increase with increasing distance from the Sun, the maximum being reached for Saturn. The linear term includes for all planets the $L_B$ contribution for the rescaling between TCB and TDB (see Eq.~\ref{Eq27}).

\begin{table*}[!ht]
 \caption{2-body contribution to the relativistic variations in $\phi$, assuming that the planets follow Keplerian orbits (see toy model of Sect.~\ref{Section31}). The second column is for the rate of the linear term. The other columns are for the amplitude of the periodic terms, following the parametrization of Eq.~(\ref{Eq2body}).}
\centering
\begin{tabular}{cccccccc}
\hline \hline
\\[-0.3cm] 
Planet & $\dot\phi_\mathrm{GR}$ (mas/day) & $\phi_{r1}$ (mas) & $\phi_{r2}$ (mas) & $\phi_{r3}$ (mas) & $\phi_{r4}$ (mas) \\
\hline
Mercury  & -0.503  &  -3.23  & -0.33  & -0.05  & -0.01  \\
Venus    & 0.026   &  0.04  & 0.00  & 0.00  & 0.00  \\
Earth    & 0.906   &  -24.85  & -0.21  & -0.00  & 0.00  \\
Mars     & 7.311   &  -166.95  & -7.78  & -0.55  & -0.05 \\
Jupiter  & 39.673  & -397.54  & -9.63  & -0.35  & -0.02 \\
Saturn   & 40.735  &  -573.00  & -15.89  & -0.66  & -0.03 \\
Uranus   & -26.584 &  418.97  & 9.69  & 0.34  & 0.01 \\
Neptune  & 29.248  & -109.94  & -0.49  & -0.00  & 0.00 \\
\hline
\label{Tab6}
\end{tabular}
\end{table*}

Table~\ref{Tab7} provides estimates for the contribution related to the Sun motion relative to the SSB, based on the toy model of Sect.~\ref{Section312}. For Each planet, seven terms at synodic periods corresponding the indirect effects of the other planets are computed. Most of these contributions have negligible amplitude. Only the giant planets induce indirect effects on the other planets larger than $0.01$ mas in amplitude, and this effect is almost zero on Mercury and Venus. The largest term (with an amplitudes above $4$ mas) applies to Saturn and is due to Jupiter.

\begin{table*}[!ht]
 \caption{Amplitudes of synodic terms (in mas) in the relativistic variations in $\phi$, related to the Sun motion relative to the SSB, based on the toy model of Sect.~\ref{Section312}. Each line corresponds to the considered body, and each row corresponds the body which indirectly acts on considered body.} 
\centering 
\begin{tabular}{ccccccccc} 
\hline \hline
& Mercury & Venus & Earth & Mars & Jupiter & Saturn & Uranus & Neptune\\
\hline
Mercury & & 0.00 & 0.00 & 0.00 & 0.00& 0.00 & 0.00 & 0.00 \\
Venus   & 0.00 & \text{} & 0.00 & 0.00 & 0.00 & 0.00 & 0.00 & 0.00 \\
Earth   & 0.00 & 0.00 & \text{} & 0.00 & 0.34 & 0.07 & 0.01 & 0.01 \\
Mars    & 0.00 & 0.00 & 0.00 & \text{} & 0.55 & 0.11 & 0.01 & 0.01 \\
Jupiter & 0.00 & 0.00 & 0.00 & 0.00 & \text{} & 1.45 & 0.11 & 0.09 \\
Saturn  & 0.00 & 0.00 & 0.00 & 0.00 & 4.52 & \text{} & 0.25 & 0.18 \\
Uranus  & 0.00 & 0.00 & 0.00 & 0.00 & 1.37 & 1.00 & \text{} & 0.38 \\
Neptune & 0.00 & 0.00 & 0.00 & 0.00 & 1.09 & 0.68 & 0.35 & \\
\hline
\label{Tab7}
\end{tabular}
\end{table*}

In Table~\ref{Tab8}, we compute the contribution related to the direct effect of the planets on each other, based on the toy model of Sect.~\ref{Section313}. For Each planet, we give the total rate of the linear term due to all the other planets, and the amplitudes of the seven individual terms at synodic periods. The linear terms are very small, as already noticed for Mars. As for the indirect terms, the synodic direct terms are mainly induced by the giant planets, with quasi zero effect for Mercury and Venus, and the largest term (with an amplitudes above $1$ mas) applies to Saturn and is due to Jupiter. 

\begin{table*}[!ht]
 \caption{Rate of the linear term and amplitudes of the synodic terms (in mas) in the relativistic variations in $\phi$, due to the direct effects of the planets, based on the toy model of Sect.~\ref{Section313}. Each line corresponds to the considered body, and each row corresponds the body which directly acts on considered body.} 
\centering 
\begin{tabular}{cccccccccc} 
\hline \hline
& $\dot\phi_\mathrm{GR}$ (mas/day) & Mercury & Venus & Earth & Mars & Jupiter & Saturn & Uranus & Neptune\\
\hline
Mercury &0.000& & 0.00 & 0.00 & 0.00 & 0.00& 0.00 & 0.00 & 0.00 \\
Venus   &0.000& 0.00 & \text{} & 0.00 & 0.00 & 0.00 & 0.00 & 0.00 & 0.00 \\
Earth   &0.003& 0.00 & 0.00 & \text{} & 0.00 &  0.03 & 0.00 & 0.00 & 0.00 \\
Mars    &0.003& 0.00 & 0.00 & 0.00 & \text{} & 0.08 & 0.01 & 0.00 & 0.00 \\
Jupiter &0.001& 0.00 & 0.00 & 0.00 & 0.00 & \text{} & 0.39 & 0.01 & 0.01 \\
Saturn  &0.003& 0.00 & 0.00 & 0.00 & 0.00 & 1.23 & \text{} & 0.06 & 0.03 \\
Uranus  &0.001& 0.00 & 0.00 & 0.00 & 0.00 & 0.17 & 0.25 & \text{} & 0.12 \\
Neptune &0.001& 0.00 & 0.00 & 0.00 & 0.00 & 0.07 & 0.10 & 0.11 & \\
\hline
\label{Tab8}
\end{tabular}
\end{table*}

The results presented in this section depend on the chosen values for the parameters, and in particular on the values of the eccentricities. We use here the eccentricities obtained from the mean orbital elements of \citet{simon:2013vz}. If necessary, the results can be refined by a fit to a numerical integration (as in Sect.~\ref{Section32}) or from a semi-analytical solution (as in Sect.~\ref{sec:VSOP}).

\section{Discussion and conclusions}

The orientation and rotation model of Mars expressed in the BCRS includes relativistic corrections. We have estimated the corrections in the Euler angles ($\psi,\varepsilon,\phi$) describing the orientation of a frame attached to the surface of Mars with respect to its mean orbit. Given the current accuracy on radioscience orbiter and lander data, a precision of $0.1$ mas in the relativistic corrections is required to avoid errors in the interpretation of measurements of Mars rotation in terms of local physics. An accurate estimation of the relativistic corrections in rotation is also useful to define IAU standards for the rotation and orientation of Mars \citep{Yse22}. 

We have considered first the relativistic terms that impact directly the rotation, and have found that only the geodetic precession induces a significant effect, and only in longitude $\psi$. Then we have investigated the terms that arise in the rotation angle $\phi$ because of the time coordinate transformation between a local Mars reference fame and the BCRS. For the longitude correction, our results are in agreement with previous results, whereas this is not the case for the rotation correction. There is no significant relativistic correction that applies to the obliquity $\varepsilon$. 

Here are our recommendations for the relativistic corrections in Mars' Euler angles (in mas):
\begin{subequations}
\begin{align}
[\psi]_\mathrm{GR}(t) &= 6.754 \, t + 0.565 \sin l'\, , \\
[\varepsilon]_\mathrm{GR}(t) &= 0\, , \\
[\phi]_\mathrm{GR}(t) &= 7.3088 \, d - 166.954 \sin l’ - 7.783 \sin 2l’ \nonumber \\
&- 0.544 \sin 3l’ + 0.567 \sin\left(\frac{2\pi}{2.235294} t + 320^\circ.997\right)\nonumber\\
&+ 0.102 \sin\left(\frac{2\pi}{2.009124} t + 303^\circ.752\right)\, , 
\label{eqfinale}
\end{align}
\end{subequations}
with $t$ and $d$ the time in years and days, respectively, and $l'$ the mean anomaly of Mars as given in Eq.~(\ref{lMa}). For the longitude angle, we keep the linear and annual terms of Eq.~(\ref{EQ13}), estimated from a Keplerian toy model, and that is consistent with the results of \citet{BMAN20}. The precision on those terms is of about $0.05\%$, as estimated from the difference with a semi-analytical derivation based on VSOP87 ephemerides (not shown here).
The linear term in $[\psi]_\mathrm{GR}$ of $6.754$ mas/yr is important, making a difference in the angle up to $135$ mas around the year 2020.
For the rotation angle, we keep the linear term and the five largest periodic terms, based on an average of the fit to the numerical solution of Table~\ref{tab:DE} and of the semi-analytical solution of Table \ref{tab:VSOP}.
The linear term in $[\phi]_\mathrm{GR}$ of $7.3$ mas/day is very large, shifting the rotation angle by $53,000$ mas around the year 2020, i.e. moving the prime meridian by almost 850 m in 20 years. The accuracy of the linear term is of the order of $10^{-5}\%$, the difference between the two solutions. The precision on the periodic terms is better than $0.01$ mas. The last two terms are at the Mars-Jupiter ($2.24$ years) and Mars-Saturn ($2.01$ years) synodic periods. Note that their phase is not the difference of the phases of the mean anomalies of Mars and Jupiter or Saturn.

Our recommendation for the expression of $[\phi]_\mathrm{GR}(t)$ replaces the estimate of \citet{YS97} (reminded in Eq.~\ref{EqYS97}), where only the three main periodic terms are given, with an error of about $9$ mas on the annual term. Such a difference can already have an effect in the radioscience data analysis, since the periods of these terms are the same as the periods of the rotation variations induced by atmosphere/surface dynamics. The synodic terms are here computed for the first time. Since their period is close the orbital period of Mars ($1.88$ year), we recommend to include them in the a priori rotation model of Mars in order to avoid any contamination of the rotation amplitudes fitted to the radio-science data. Not taking them into account would likely affect the estimate of the annual term in $\phi$ by $0.6$ mas and $0.1$ mas, respectively (if the annual term absorbs their full contribution).

The methods (analytical, numerical, and semi-analytical) presented in this study can be extended to other bodies orbiting the Sun. In Sect.~\ref{Section5.2}, we have already applied the analytical toy model to the other planets of the solar system. These results can be refined by a fit to a numerical integration or from a semi-analytical solution. The methods should also be upgraded to moons like the Galilean moons.

\begin{acknowledgements} We thank the editor and an anonymous reviewer for their careful reading and comments that have helped to improve our paper.
This work was financially supported by the Belgian PRODEX program managed 
by the European Space Agency in collaboration with the Belgian Federal 
Science Policy Office. 
This is InSight contribution ICN 304.
\end{acknowledgements}

%
%

\bibliographystyle{aa} 
\bibliography{article.bib} 

\end{document}